\documentclass[11pt]{article}
\usepackage[margin=1in]{geometry}
\usepackage[T1]{fontenc}
\usepackage{times}
\usepackage{natbib}
\usepackage{amsmath,amssymb,amsthm,mathtools,booktabs,multirow,graphicx,float}
\usepackage{tikz}
\usetikzlibrary{patterns,positioning,decorations.pathreplacing,calc,arrows.meta,fit,backgrounds}
\definecolor{okgreen}{RGB}{40,120,60}
\definecolor{caublue}{RGB}{40,90,180}
\definecolor{badred}{RGB}{200,50,50}
\definecolor{cellgray}{RGB}{110,110,110}
\definecolor{morange}{RGB}{230,126,34}
\usepackage{hyperref}
\hypersetup{colorlinks=true,linkcolor=blue!45!black,citecolor=blue!45!black,urlcolor=blue!45!black}
\usepackage{url}

\theoremstyle{plain}
\newtheorem{theorem}{Theorem}[section]
\newtheorem{proposition}[theorem]{Proposition}
\newtheorem{corollary}[theorem]{Corollary}
\newtheorem{lemma}[theorem]{Lemma}
\newtheorem{counterexample}[theorem]{Counterexample}
\newtheorem{assumption}[theorem]{Assumption}
\newtheorem{remark}[theorem]{Remark}
\theoremstyle{definition}
\newtheorem{definition}[theorem]{Definition}
\theoremstyle{plain}
\newtheorem{innercustomthm}{Theorem}
\newenvironment{customthm}[1]{\renewcommand\theinnercustomthm{#1}\innercustomthm}{\endinnercustomthm}
\newtheorem{innercustomlem}{Lemma}
\newenvironment{customlem}[1]{\renewcommand\theinnercustomlem{#1}\innercustomlem}{\endinnercustomlem}
\newtheorem{innercustomcor}{Corollary}
\newenvironment{customcor}[1]{\renewcommand\theinnercustomcor{#1}\innercustomcor}{\endinnercustomcor}

\DeclareMathOperator{\Tr}{tr}
\DeclareMathOperator{\rank}{rank}

\DeclareMathOperator{\ran}{ran}
\newcommand{\C}{\mathbb{C}}
\newcommand{\R}{\mathbb{R}}
\newcommand{\A}{\mathcal{A}}
\newcommand{\Herm}{\mathbb{H}}
\newcommand{\Fro}[1]{\lVert #1\rVert_{F}}
\newcommand{\Sph}{\mathbb{S}}
\newcommand{\Pc}{\mathcal{P}}
\newcommand{\Zc}{\mathcal{Z}}
\newcommand{\Dc}{\mathcal{D}}

\title{Rank-one Factored Gradient Descent without Restricted Isometry: 
Application to Quantum Pure State Tomography on Multi-Qubits}
\author{Takahiro Kajisa\\[2pt]
\normalsize The University of Electro-Communications, Chofu, Tokyo, Japan\\
\normalsize\texttt{t.kajisa@uec.ac.jp}}
\date{}

\begin{document}
\maketitle
\begin{abstract}
Factored gradient descent (FGD) is a scalable nonconvex method for low-rank positive semidefinite matrix recovery, in which the rank and positivity constraints are satisfied automatically by the factorization. Most of the existing theory that certifies benignity of its loss landscape, however, relies on a restricted isometry property (RIP) of the measurement map. In quantum state tomography, one of its applications, the measurements that can be acquired with stable statistics are limited by detection efficiency and drift to Pauli observables of low weight; we first show that under such partial-weight Pauli measurements RIP fails in the worst possible way. We then bundle the mutually commuting observables setting by setting, pushing the state forward to the probability distribution of each setting, and introduce a max-sliced total-variation loss. On the stratum where the setting attaining the maximum is unique there is no spurious local minimum, and provided only that the measurement map is injective on generic pure states, the global minimum is the truth up to a phase. Finally we verify these claims numerically on four qubits and compare against FGD with the ordinary loss.
\end{abstract}

\section{Introduction}\label{sec:intro}

Write $\Herm_{d}$ for the space of Hermitian matrices. The basic problem of low-rank matrix recovery is to recover an unknown positive semidefinite matrix $M_{\star}\in\Herm_{d}$ of rank $r\ll d$ from the observation $y=\A(M_{\star})$, where the linear measurement map $\A:\Herm_{d}\to\R^{m_{0}}$, $\A(M)=\bigl(\Tr(A_{1}M),\dots,\Tr(A_{m_{0}}M)\bigr)$, is determined by known measurement matrices $A_{1},\dots,A_{m_{0}}\in\Herm_{d}$. Burer--Monteiro type methods factor the optimization variable as $M=YY^{\dagger}$ and solve the nonconvex least-squares problem
\begin{equation}\label{eq:fgd}
\min_{Y\in\C^{d\times r}}\ f_{r}(Y)=\tfrac12\bigl\lVert\A(YY^{\dagger})-y\bigr\rVert_{2}^{2}
\end{equation}
by gradient methods (factored gradient descent, FGD) \citep{burer2003local}. The factorization makes the positivity and the rank constraint on $M$ automatic and reduces the number of parameters from $d^{2}$ to $dr$. To overcome the nonconvexity that this introduces, it is standard to impose a restricted isometry property (RIP) on the measurement map: $\A$ satisfies rank-$r$ RIP if there is $\delta_{r}\in[0,1)$ such that
\begin{equation}\label{eq:rip}
(1-\delta_r)\,\Fro{X}^{2}\ \le\ \lVert\A(X)\rVert_{2}^{2}\ \le\ (1+\delta_r)\,\Fro{X}^{2}
\end{equation}
for every $X\in\Herm_{d}$ of rank at most $r$; the smallest such value is the RIP constant $\delta_{r}$. If $\delta_{2r}$ is small enough, the landscape of \eqref{eq:fgd} is benign: there is no spurious local minimum and every saddle point is strict \citep{bhojanapalli2016global,ge2017no,zhu2018global,chi2019nonconvex}. For rank $r=1$ even the location of the threshold is known. \citet{zhang2019sharp} proved that $\delta_{2}<1/2$ excludes spurious local minima, and at the same time constructed a measurement map and a ground truth with $\delta_{2}=1/2$ that does have one. By monotonicity of the RIP constant such a pair exists for every $\delta_{2}\ge1/2$, so $1/2$ is sharp in the worst-case sense over the class of measurement maps.

Quantum state tomography (QST) is a special case of this framework. The state of an $n$-qubit system is a positive semidefinite matrix $M\in\Herm_{d}$ ($d=2^{n}$) with $\Tr M=1$ on the $n$-fold tensor product $(\C^{2})^{\otimes n}$, and measuring an observable $A_{i}\in\Herm_{d}$ on $M$ means obtaining $\Tr(A_{i}M)$. Choosing the observables $A_{1},\dots,A_{m}$ is therefore nothing but choosing the measurement map $\A$. In the noiseless idealization, QST can thus be written as the feasibility problem of finding a state consistent with the measured values $y=\A(M_{\star})$:
\begin{equation}\label{eq:qst}
\begin{aligned}
&\text{find} && M\in\Herm_{d}\\
&\text{subject to} && M\succeq 0,\qquad \Tr M=1,\qquad
\Tr(A_{i}M)=y_{i}\quad(i=1,\dots,m).
\end{aligned}
\end{equation}
When the measurement determines a rank-$r$ state uniquely among all states, the feasible set collapses to the single point $\{M_{\star}\}$ and solving this problem is recovery itself \citep{kalev2015quantum,baldwin2016strictly}. Moreover, what a quantum experiment aims to prepare is in most cases a pure state---a density matrix of rank $1$ \citep{eisert2020quantum}. The convex formulation \eqref{eq:qst} cannot exploit this prior knowledge and does not guarantee that its solution has rank $1$. If one wants a pure state as the output, the natural choice is the Burer--Monteiro factorization, which builds positivity and rank into the search space from the start \citep{kyrillidis2018provable,kim2023fast}. Indeed, replacing the equality constraints of \eqref{eq:qst} by a residual sum of squares, restricting the variable to rank-one matrices and imposing the normalization gives the spherical version of \eqref{eq:fgd},
\begin{equation}\label{eq:fgd-sphere}
\min_{Y\in \mathbb{S}_{d}}\ \tfrac12\bigl\lVert\A(YY^{\dagger})-y\bigr\rVert_{2}^{2},\qquad \mathbb{S}_{d}=\{\,u\in\mathbb{C}^{d}\mid \lVert u\rVert=1\,\}.
\end{equation}
Analyzing the landscape of this nonconvex formulation \eqref{eq:fgd-sphere} of pure-state QST is the main purpose of this paper.

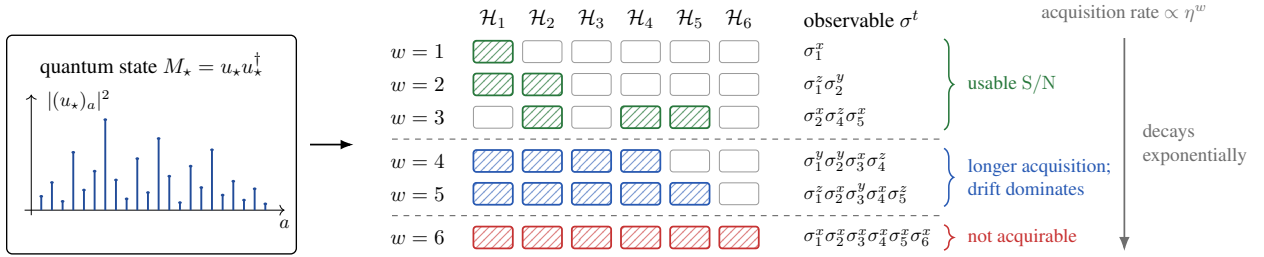
\begin{figure}[htbp]
\centering
\resizebox{\linewidth}{!}{%
\begin{tikzpicture}[
  x=1cm,y=1cm,
  cell/.style={rounded corners=1.6pt,minimum width=0.72cm,minimum height=0.40cm,
               draw=cellgray,line width=0.5pt,inner sep=0pt},
  actgreen/.style={cell,draw=okgreen,line width=0.9pt,pattern=north east lines,pattern color=okgreen!70},
  actblue/.style={cell,draw=caublue,line width=0.9pt,pattern=north east lines,pattern color=caublue!60},
  actred/.style={cell,draw=badred,line width=0.9pt,pattern=north east lines,pattern color=badred!60},
  idle/.style={cell,draw=cellgray!70},
]

\draw[rounded corners=3pt,line width=0.7pt] (-8.9,0.85) rectangle (-3.6,-3.15);
\node[anchor=north,font=\normalsize] at (-6.25,0.68) {quantum state $M_{\star}=u_{\star}u_{\star}^{\dagger}$};
\begin{scope}[shift={(-8.45,-2.35)},x=0.61cm,y=0.46cm]
  \draw[->,line width=0.5pt] (-0.2,0) -- (7.6,0);
  \draw[->,line width=0.5pt] (0,-0.2) -- (0,4.6);
  \foreach \x/\h in {0.30/0.55, 0.62/1.10, 0.94/0.35, 1.26/2.30, 1.58/0.80,
                     1.90/1.55, 2.22/3.60, 2.54/1.20, 2.86/0.45, 3.18/2.05,
                     3.50/0.70, 3.82/2.85, 4.14/1.35, 4.46/0.30, 4.78/1.75,
                     5.10/0.90, 5.42/2.40, 5.74/0.60, 6.06/1.15, 6.38/0.40,
                     6.70/0.85, 7.02/0.25}{
    \draw[caublue!85!black,line width=1.0pt] (\x,0) -- (\x,\h);
    \fill[caublue!85!black] (\x,\h) circle (0.055);
  }
  \node[font=\small,anchor=north] at (7.6,-0.1) {$a$};
  \node[font=\small,anchor=west] at (0.25,4.35) {$|(u_{\star})_{a}|^{2}$};
\end{scope}

\draw[-{Latex[length=2.4mm]},line width=0.9pt] (-3.35,-1.15) -- (-2.55,-1.15);

\foreach \c/\lab in {1/{1},2/{2},3/{3},4/{4},5/{5},6/{6}}{
  \node[font=\normalsize] at (\c*0.9-0.9,1.15) {$\mathcal{H}_{\lab}$};
}
\node[font=\normalsize,anchor=west] at (5.55,1.15) {observable $\sigma^{t}$};

\newcommand{\drawrow}[5]{%
  \foreach \c in {1,...,6}{
    \node[idle] at (\c*0.9-0.9,#1) {};
  }
  \foreach \c in {#2}{
    \node[#3] at (\c*0.9-0.9,#1) {};
  }
  \node[font=\normalsize,anchor=east] at (-0.75,#1) {#4};
  \node[font=\small,anchor=west] at (5.55,#1) {#5};
}

\drawrow{0.55}{1}{actgreen}{$w=1$}{$\sigma^{x}_{1}$}
\drawrow{-0.05}{1,2}{actgreen}{$w=2$}{$\sigma^{z}_{1}\sigma^{y}_{2}$}
\drawrow{-0.65}{2,4,5}{actgreen}{$w=3$}{$\sigma^{x}_{2}\sigma^{z}_{4}\sigma^{x}_{5}$}
\drawrow{-1.45}{1,2,3,4}{actblue}{$w=4$}{$\sigma^{y}_{1}\sigma^{y}_{2}\sigma^{x}_{3}\sigma^{z}_{4}$}
\drawrow{-2.05}{1,2,3,4,5}{actblue}{$w=5$}{$\sigma^{z}_{1}\sigma^{x}_{2}\sigma^{y}_{3}\sigma^{x}_{4}\sigma^{z}_{5}$}
\drawrow{-2.85}{1,2,3,4,5,6}{actred}{$w=6$}{$\sigma^{x}_{1}\sigma^{x}_{2}\sigma^{x}_{3}\sigma^{x}_{4}\sigma^{x}_{5}\sigma^{x}_{6}$}

\draw[dashed,cellgray,line width=0.6pt] (-1.85,-1.05) -- (8.25,-1.05);
\draw[dashed,cellgray,line width=0.6pt] (-1.85,-2.45) -- (8.25,-2.45);

\draw[decorate,decoration={brace,amplitude=5pt,mirror},okgreen,line width=0.8pt]
  (8.25,-0.88) -- (8.25,0.78);
\node[okgreen,font=\small,align=left,anchor=west] at (8.55,-0.05)
  {usable $\mathrm{S/N}$};

\draw[decorate,decoration={brace,amplitude=5pt,mirror},caublue,line width=0.8pt]
  (8.25,-2.28) -- (8.25,-1.22);
\node[caublue,font=\small,align=left,anchor=west] at (8.55,-1.75)
  {longer acquisition;\\ drift dominates};

\draw[decorate,decoration={brace,amplitude=5pt,mirror},badred,line width=0.8pt]
  (8.25,-3.08) -- (8.25,-2.62);
\node[badred,font=\small,align=left,anchor=west] at (8.55,-2.85)
  {not acquirable};

\draw[-{Latex[length=2.6mm]},line width=0.9pt,cellgray] (11.55,0.80) -- (11.55,-3.10);
\node[font=\small,anchor=south,cellgray] at (11.55,0.95) {acquisition rate $\propto\eta^{w}$};
\node[font=\small,anchor=west,cellgray,align=left] at (11.75,-1.15)
  {decays\\ exponentially};

\end{tikzpicture}%
}
\vspace*{-6mm}
\caption{Partial-weight Pauli measurements. Each row is one observable $\sigma^{t}$; the shaded boxes mark the sites on which the string acts nontrivially, and the weight $w(t)$ is the number of such sites. Measuring an observable of weight $w$ requires reading out $w$ sites jointly, so the acquisition rate decays exponentially as $\eta^{w}$, where $\eta\in(0,1)$ is the per-site detection efficiency. Accumulating high-weight statistics therefore takes a long time, during which the apparatus drifts and the state itself changes---the samples cease to be i.i.d.\ and a systematic bias enters. The coherence that distinguishes the family $\psi_{\varphi}$ of Counterexample~\ref{cex:ghz} appears only in the $w=n$ string, which falls outside the usable regime.}
\label{fig:weight}
\end{figure}

Nevertheless, FGD has not become the mainstream tool in QST. One major reason is that most theorems certifying benignity of its landscape or escape from saddle points rest on RIP of the measurement map \citep{bhojanapalli2016global,kyrillidis2018provable}. Realizing a RIP measurement system in an actual quantum experiment is difficult, and the assumption should not be used uncritically. The random Pauli measurements for which RIP is known to hold \citep{liu2011universal}, for instance, require building the measurement pool uniformly at random under the premise that all $n$ qubits can be read out simultaneously; in practice the success rate of such a joint readout decays exponentially in $n$. This is not a quirk of a particular device but a statistical consequence of incorporating imperfect detection efficiency into the model. And if one lengthens the acquisition time in order to collect such full-qubit coincidence data, drift of every device from state preparation to readout becomes unavoidable, so in principle only partial joint-readout events can be used as data (Figure~\ref{fig:weight}). The situation is different for the tomography of pure states. It is known that almost every pure state is determined uniquely, among all pure states, by the Pauli expectations of subsystems of roughly half the size alone \citep{linden2002almost,linden2002parts,jones2005parts,chen2013uniqueness}. Moreover, since the search space of rank-$1$ FGD is the manifold of pure states itself, for the global solution to be confined to the truth it suffices that the measurement map be generically injective on the pure-state manifold. That is, as long as pure states are the target, high-weight Pauli measurements spanning all qubits are unnecessary for identifiability. In this paper we adopt the explicit construction of \citet{zsz2024}---the Pauli expectations of weight at most $N$ on a $2N$-qubit system determine a generic pure state uniquely---and work with this partial measurement map. What is obtained in this way, however, is identifiability and not RIP. Uniqueness is a statement about ``almost every'' pure state and allows an exceptional set. Indeed, the following GHZ-type family is not determined by partial measurements even among pure states \citep{walck2008only}. RIP, on the other hand, is a uniform condition over all rank-$2$ matrices that does not depend on the truth, so a single such exceptional family breaks it in the worst possible way, $\delta_{2}\ge1$, however generic the truth may be. The following counterexample is exactly this.

\begin{counterexample}[Counterexample to RIP with $\delta_{2}\ge1$]\label{cex:ghz}
Let $e_{0},e_{1}$ be the standard basis of $\C^{2}$ and put $\psi_{\varphi}=\tfrac{1}{\sqrt2}\bigl(e_{0}^{\otimes n}+e^{\mathrm{i}\varphi}\,e_{1}^{\otimes n}\bigr)$ for $\varphi\in[0,2\pi)$. For $\varphi\neq0$ this is a pure state different from $\psi_{0}$. Yet on any proper subset $S\subsetneq\{1,\dots,n\}$ the reduced density matrix
\[
\rho^{S}:=\Tr_{S^{c}}\bigl(\psi_{\varphi}\psi_{\varphi}^{\dagger}\bigr)
\;=\;\tfrac12\Bigl((e_{0}e_{0}^{\dagger})^{\otimes S}\;+\;(e_{1}e_{1}^{\dagger})^{\otimes S}\Bigr)
\]
takes a value independent of $\varphi$. Hence for every measurement of the form $A=A^{S}\otimes I^{S^{c}}$ with $\lvert S\rvert=N<n$ the expectation is $\Tr(A\,\psi_{\varphi}\psi_{\varphi}^{\dagger})=\Tr\bigl(A^{S}\rho^{S}\bigr)$, so $\psi_{0}$ and $\psi_{\varphi}$ cannot be distinguished. Consequently the rank-two Hermitian matrix $X_{\varphi}=\psi_{0}\psi_{0}^{\dagger}-\psi_{\varphi}\psi_{\varphi}^{\dagger}$ determined by the difference of the two pure states lies in the kernel of any measurement map $\A$ built solely from the partial measurements $\{A^{S}\otimes I^{S^{c}}\mid A^{S}\in\Herm_{2^{\lvert S\rvert}}\}_{S\subsetneq\{1,\dots,n\}}$. The lower bound in \eqref{eq:rip} therefore drops to $0$ and $\delta_{2}\ge1$, i.e.\ RIP does not hold.
\end{counterexample}

The starting point of this work is to look squarely at this structural mismatch between the design of the measurement and the design of the loss. The $\ell^{2}$ loss demands, as a regularity condition on the landscape, information about high-order coherences that a partial measurement family cannot touch even in principle. We therefore rebuild the $\ell^{2}$ geometry so that the measurement design and the regularity condition are reconciled. The key is that within one measurement setting the observables commute. Fixing a setting $\tau$, the observables that can be read out simultaneously are diagonalized by a common family of eigenprojectors $\{\Pi^{\tau}_{k}\}_{k}$ with $\sum_{k}\Pi^{\tau}_{k}=I$, and every one of them can be written as $A=\sum_{k}a_{k}\Pi^{\tau}_{k}$. The measurement map then reads
\[
\Tr\bigl(A\,uu^{\dagger}\bigr)\;=\;\sum_{k}a_{k}\,p^{\tau}_{k}(u),
\quad
p^{\tau}_{k}(u)\;=\;u^{\dagger}\Pi^{\tau}_{k}u,
\]
so it depends on the state only through the probability vector $p^{\tau}(u)=(p^{\tau}_{k}(u))_{k}$. In other words, one setting is a map $u\mapsto p^{\tau}(u)$ that pushes the state forward to a finitely supported probability distribution. If we increase the number of settings $\tau$ and bundle the observables setting by setting into slices, noncommutativity is dissolved inside each slice and what remains is the classical problem of comparing two probability vectors. Writing $u_{\star}u_{\star}^{\dagger}$ for the pure state to be recovered and $q^{\tau}:=p^{\tau}(u_{\star})$ for its pushforward, we take as the loss the worst-case total variation between the distributions $p^{\tau}(u)$ and $q^{\tau}$ over all slices,
\[
L(u)\;=\;\max_{\tau}\;\bigl\lVert p^{\tau}(u)-q^{\tau}\bigr\rVert_{1}.
\]
This is an analogue of the max-sliced Wasserstein metric, except that the variable selecting the slice is not a free angle but the experimental setting itself, and the distance on each slice is total variation. The price is nonsmoothness; it is a price worth paying, and it is precisely that nonsmoothness which makes the analysis possible. Our main theorem takes the following form.

\begin{theorem}[Main theorem, informal]
On the set of points $u$ with $L(u)>0$ at which the slice attaining the maximum in $L$ is unique, every stationary point is a saddle point, and an escape direction is constructed explicitly. Moreover, if the data are exact and the measurement map $\A$ determines $u_{\star}$ uniquely among pure states, then $L(u)=0$ if and only if $u=e^{\mathrm{i}\theta}u_{\star}$ for some $\theta\in\R$.
\end{theorem}

Our numerical experiments use $4$ qubits ($N=2$, pair blocks). Comparing on identical data from identical random initial points, in the noiseless case the subgradient method (with momentum) on the max-sliced loss recovers the truth in $100/100$ trials, whereas the ordinary $\ell^{2}$ loss is trapped at a spurious local minimum in $25$ of them. With finite shots, a two-stage scheme---descend on the max-sliced loss, then polish with the $\ell^{2}$ loss---is best at every shot budget we tried.

\begin{remark}[Related work: matrix completion]
Even for ordinary FGD there are situations in which benignity is recovered without RIP, the prime example being matrix completion. There the measurement map extracts entries, and it annihilates the matrix $e_{i}e_{j}^{\top}+e_{j}e_{i}^{\top}$ supported on a single unobserved entry $(i,j)$, so RIP fails with $\delta_{2}\ge1$. Even so, \citet{ge2016matrix,ge2017no} assumed that the rows of the truth have uniformly small norms (incoherence) and added a penalty on factors with large row norms, and proved the absence of spurious local minima. The mechanism is that the kernel consists of matrices whose mass is concentrated on a few entries, so one keeps the search away from such matrices and recovers an isometry there. The state of Counterexample~\ref{cex:ghz} is likewise localized on a few components, but we regard even a neighbourhood of it as an object to be recovered, so such a keep-away prescription is unavailable here.
\end{remark}

\begin{remark}[Related work: convex relaxations for low-rank tomography]
Low-rank state tomography was originally formulated as compressed sensing: $O(rd\,\mathrm{polylog}\,d)$ random Pauli expectations determine a rank-$r$ state, and under RIP-type conditions convex programs recover it stably \citep{gross2010quantum,liu2011universal,flammia2012quantum,kueng2017low,kalev2015quantum,baldwin2016strictly}. Under noise, however, the rank of the convex solution is a by-product of the noise and the active constraints and generally differs from $r$ \citep{blumekohout2010hedged,scholten2018behavior}, so a rank-$r$ description requires an a posteriori truncation, after which the estimate is no longer the solution of the convex program. Factorization-based methods avoid this by optimizing inside the model class \citep{kyrillidis2018provable,hsu2024quantum}. Moreover, our partial measurements do not provide the uniqueness among all states that the convex guarantees presuppose---only generic injectivity on the pure-state manifold---so recovering pure states here is effectively limited to nonconvex formulations.
\end{remark}

\section{Preliminaries}\label{sec:setup}
\subsection{Quantum state tomography on $2N$ qubits}\label{ssec:qst}
The state space of a qubit, the minimal unit of quantum information, is $\C^{2}$, and a quantum state on it---a positive semidefinite matrix $M\in\Herm_{2}$ with $\Tr M=1$---can be written uniquely as
\[
M\;=\;\tfrac12\bigl(I+r_{x}\sigma^{x}+r_{y}\sigma^{y}+r_{z}\sigma^{z}\bigr),
\qquad r_{a}=\Tr(\sigma^{a}M)\in\R,
\]
\begin{equation}\label{eq:pauli}
\sigma^{x}=\begin{pmatrix}0&1\\1&0\end{pmatrix},\qquad
\sigma^{y}=\begin{pmatrix}0&-\mathrm{i}\\\mathrm{i}&0\end{pmatrix},\qquad
\sigma^{z}=\begin{pmatrix}1&0\\0&-1\end{pmatrix},\qquad \sigma^{0}=I .
\end{equation}
This is the Bloch vector expansion of $M$. Here $M\succeq0$ is equivalent to $\lVert r\rVert\le1$, and purity $\rank M=1$ to $\lVert r\rVert=1$. The three matrices $\sigma^{x},\sigma^{y},\sigma^{z}$ appearing as coefficients are the Pauli matrices, and each component of the Bloch vector is nothing but their expectation: knowing the state of a qubit is the same as knowing the expectations of the three Pauli matrices. Expectations, however, are not directly readable quantities, and this is where the quantum measurement procedure enters. Each $\sigma^{a}$ ($a\in\{x,y,z\}$) has eigenvalues $\pm1$; writing $v^{a}_{k}$ for the unit eigenvector belonging to the eigenvalue $(-1)^{k}$ ($k\in\{0,1\}$) and
\[
\Pi^{a}_{k}:=v^{a}_{k}(v^{a}_{k})^{\dagger}=\tfrac12\bigl(I+(-1)^{k}\sigma^{a}\bigr)
\]
for the corresponding projector, the condition $\sum_{k=0,1}\Pi^{a}_{k}=I$ of a projective measurement holds. Looking at the expectation of $\sigma^{a}$ then means reading the state out repeatedly with detectors corresponding to the two projectors $\{\Pi^{a}_{0},\Pi^{a}_{1}\}$. One readout returns a single $k\in\{0,1\}$, with probability given by the Born rule
\[
\mathbb{P}[k]\;=\;\Tr\bigl(\Pi^{a}_{k}M\bigr)\;=\;\tfrac12\bigl(1+(-1)^{k}r_{a}\bigr)
\]
(and $\sum_{k}\mathbb{P}[k]=1$ follows from $\sum_{k}\Pi^{a}_{k}=I$). Hence $R$ readouts are i.i.d.\ Bernoulli trials and the number $R_{0}$ of outcomes $k=0$ follows the binomial distribution
\[
\mathbb{P}[R_{0}=s]\;=\;\binom{R}{s}\Bigl(\frac{1+r_{a}}{2}\Bigr)^{\!s}\Bigl(\frac{1-r_{a}}{2}\Bigr)^{\!R-s}.
\]
The estimator built from the relative frequency is $\hat r_{a}=2R_{0}/R-1$, with $\mathbb{E}[\hat r_{a}]=r_{a}$ and $\mathrm{Var}[\hat r_{a}]=(1-r_{a}^{2})/R$. The primary data of an experiment are thus counts rather than expectations, and the estimate $\hat r$ built from them can easily violate, at finite sample size, not only the purity condition $\lVert\hat r\rVert=1$ but even the condition $\lVert\hat r\rVert\le1$ of being a quantum state at all. This shows how useful it is to impose prior knowledge as a hard constraint on the optimization variable.

An $n$-qubit system is described by the tensor product: the state space is $(\C^{2})^{\otimes n}\cong\C^{d}$ with $d=2^{n}$, and a state is a positive semidefinite $M\in\Herm_{d}$ with $\Tr M=1$. A state with $\rank M=1$ is called pure and is written $M=uu^{\dagger}$ with $u\in\Sph_{d}$, the vector $u$ being unique up to a global phase. The Pauli strings
\[
P_{b}\;=\;\sigma^{b_{1}}\otimes\cdots\otimes\sigma^{b_{n}},\qquad b=(b_{1},\dots,b_{n})\in\{x,y,z,0\}^{n},
\]
satisfy $\Tr(P_{b}P_{b'})=d\,\delta_{bb'}$ and form an orthogonal basis of $\Herm_{d}$, and the Bloch expansion generalizes to
\begin{equation}\label{eq:bloch}
M\;=\;\frac{1}{d}\sum_{b\in\{x,y,z,0\}^{n}}\left[\,\Tr(P_{b}M)\,\right]\,P_{b}.
\end{equation}
Since $\Tr(IM)=1$ is the normalization, a state is determined by the remaining $4^{n}-1$ Pauli expectations. The number of qubits on which $P_{b}$ acts nontrivially, i.e.\ the number of indices $i$ with $b_{i}\neq 0$, is called the weight of $P_{b}$ and written $w(b)$. The measurement procedure is inherited from the single-qubit case. Choosing a readout axis $b_{i}\in\{x,y,z\}$ at each site, the product basis $e_{b,k}=\bigotimes_{i}v^{b_{i}}_{k_{i}}$ ($k\in\{0,1\}^{n}$) is an orthonormal basis of $\C^{d}$, and the projectors
\[
\Pi^{b}_{k}:=e_{b,k}e_{b,k}^{\dagger},\qquad \sum_{k\in\{0,1\}^{n}}\Pi^{b}_{k}=I
\]
define a projective measurement. One readout returns a single bit string $k$, with probability $\mathbb{P}[k]=\Tr(\Pi^{b}_{k}M)$ by the axioms, and the counts over $R$ readouts follow, instead of a binomial, the multinomial distribution $\mathrm{Mult}\bigl(R,(\Tr(\Pi^{b}_{k}M))_{k}\bigr)$ over the $2^{n}$ cells. What differs from the single-qubit case is that one readout yields several observables at once. Expanding the tensor product of $\Pi^{a}_{k}=\tfrac12\bigl(I+(-1)^{k}\sigma^{a}\bigr)$ gives
\begin{equation}\label{eq:substring}
\Pi^{b}_{k}\;=\;2^{-n}\sum_{T\subseteq\{1,\dots,n\}}(-1)^{\sum_{i\in T}k_{i}}\,P_{b,T},
\qquad
P_{b,T}=\bigotimes_{i\in T}\sigma^{b_{i}},
\end{equation}
and conversely, by orthogonality of the Walsh functions $\sum_{k}(-1)^{\sum_{i\in T}k_{i}+\sum_{i\in T'}k_{i}}=2^{n}\delta_{T,T'}$, multiplying both sides of \eqref{eq:substring} by $(-1)^{\sum_{i\in T'}k_{i}}$ and summing over $k$ leaves only the term $T=T'$, so that
\begin{equation}\label{eq:substring_inverse}
P_{b,T}=\sum_{k}(-1)^{\sum_{i\in T}k_{i}}\Pi^{b}_{k}.
\end{equation}
Thus choosing one axis assignment $b$ yields simultaneously the expectations of all $2^{n}$ Pauli strings $\{P_{b,T}\}_{T}$ along $b$, as signed sums of the same counts; conversely the expectation of a string not along $b$ cannot be obtained from that readout. Designing a measurement thus amounts to choosing which axis assignments $b$ to prepare, and how many. Measuring a string of weight $w$, however, requires reading out $w$ sites jointly, so with per-site efficiency $\eta\in(0,1)$ the acquisition rate decays exponentially as $\eta^{w}$. Moreover, for large $n$ the device drift sets in before high-weight statistics can be accumulated well above the noise, so what is used in the laboratory are measurements restricted to low weight (Figure~\ref{fig:weight}). We therefore adopt the following standing assumption.

\begin{assumption}[Measurement model]\label{as:meas}
In an $n=2N$ qubit system, the object of measurement is a pure state $M_{\star}=u_{\star}u_{\star}^{\dagger}$, and the expectations of Pauli strings up to weight $N$ are assumed to be statistically acquirable.
\end{assumption}

On the other hand, it is known that almost every pure state of a $2N$-qubit system is uniquely determined by its $N$-body reduced density matrices, i.e.\ by the Pauli expectations of weight at most $N$. Partial-weight Pauli measurements are thus a design forced by experimental constraints and justified theoretically at the same time.

\begin{proposition}[\citep{zsz2024}]\label{prop:udp}
Under the measurements of Assumption~\ref{as:meas}, almost every pure state is uniquely determined within the set of all pure states.
\end{proposition}

\subsection{Max-sliced loss function}\label{ssec:loss}

Based on Assumption~\ref{as:meas} we now define the loss used in this paper, after fixing some terminology.

\begin{definition}[Setting and slice]\label{def:slice}
Write $v^{a}_{k}$ for the unit eigenvector of $\sigma^{a}$ ($a\in\{x,y,z\}$) belonging to the eigenvalue $(-1)^{k}$, $k\in\{0,1\}$. A \emph{measurement setting} is a pair $\tau=(S,b)$ of a subset of qubits $S\subset\{1,2,\dots,2N\}$ with $|S|=N$ and a label string $b\in\{x,y,z\}^{S}$. The measurement in the setting $\tau$ is described, through the orthonormal basis $\{e_{b,k}=\bigotimes_{i\in S}v^{b_{i}}_{k_{i}}:k\in\{0,1\}^{S}\}$ of the $S$-qubits, by
\begin{equation}\label{eq:resolution}
\Pi^{\tau}_{k}\;=\;e_{b,k}e_{b,k}^{\dagger}\otimes I ,\qquad \sum_{k}\Pi^{\tau}_{k}=I .
\end{equation}
For fixed $\tau$ the $\{\Pi^{\tau}_{k}\}_{k}$ are mutually orthogonal projectors forming a resolution of the identity $\sum_{k}\Pi^{\tau}_{k}=I$. What can be read out simultaneously in the setting $\tau$ is the family of all Pauli substrings diagonal in the readout basis, hence mutually commuting. We call them a \emph{slice} (Figure~\ref{fig:slice}(a)).
\end{definition}

\begin{definition}[Cell distribution]
When the measured object is a pure state $M=uu^{\dagger}\in\Herm_{d}$, the probability that one measurement returns the bit string $k$ is
\begin{equation}\label{eq:born}
p^{\tau}_{k}(u)\;=\;u^{\dagger}\,\Pi^{\tau}_{k}\,u
\end{equation}
by the axioms of quantum mechanics. We call the realized value $k$ a \emph{cell}, $p^{\tau}(u)=(p^{\tau}_{k}(u))_{k}$ the \emph{cell distribution} of the setting $\tau$, and, writing $q^{\tau}_{k}$ for the relative frequency of the cell $k$, $q^{\tau}=(q^{\tau}_{k})_{k}$ the \emph{empirical cell distribution}.
\end{definition}

\begin{definition}[Max-sliced loss]\label{def:loss}
For a model $M=uu^{\dagger}$ with $u\in\Sph_{d}$, let $p^{\tau}(u)=(p^{\tau}_{k}(u))_{k}$ be the cell probabilities in the setting $\tau$, let $q^{\tau}=(q^{\tau}_{k})_{k}$ be the empirical cell distribution obtained from the truth $M_{\star}=u_{\star}u_{\star}^{\dagger}$, and write $\delta^{\tau}_{k}(u)=p^{\tau}_{k}(u)-q^{\tau}_{k}$ for the residual inside the setting $\tau$. The total-variation loss of the slice $\tau$ and the max-sliced loss $L$ are defined by 
\begin{equation}\label{eq:sliceloss}
\ell_{\tau}(u):=\;\bigl\lVert p^{\tau}(u)-q^{\tau}\bigr\rVert_{1}\;=\;\sum_{k}\bigl\lvert\delta^{\tau}_{k}(u)\bigr\rvert,
\end{equation}
\begin{equation}\label{eq:loss}
L(u):=\;\max_{\tau}\;\ell_{\tau}(u).
\end{equation}
\end{definition}

Directly from the definition, mass is conserved in every slice:
\begin{equation}\label{eq:mass}
\sum_{k}\delta^{\tau}_{k}(u)\;=\;0.
\end{equation}
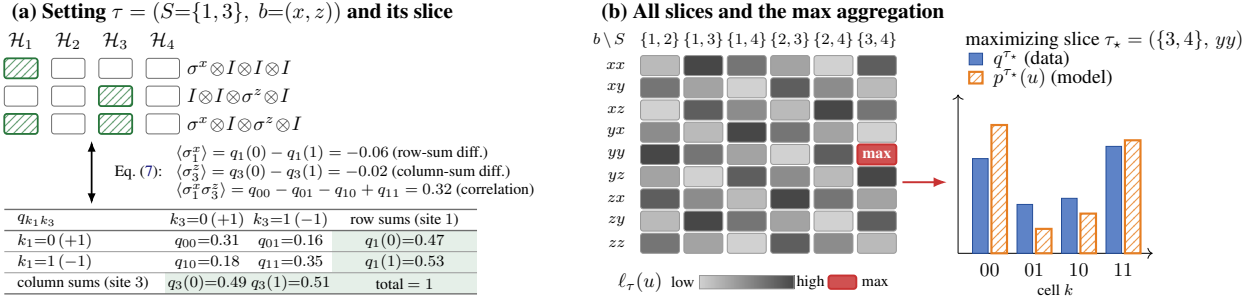
\begin{figure}[t]
\centering
\resizebox{\linewidth}{!}{%
\begin{tikzpicture}[
  x=1cm,y=1cm,
  cell/.style={rounded corners=1.4pt,minimum width=0.6cm,minimum height=0.36cm,
               draw=cellgray,line width=0.5pt,inner sep=0pt},
  act/.style={cell,draw=okgreen,line width=0.9pt,pattern=north east lines,pattern color=okgreen!70},
]
\node[font=\normalsize\bfseries,anchor=west] at (-0.3,3.35) {(a) Setting $\tau=(S{=}\{1,3\},\ b{=}(x,z))$ and its slice};
\foreach \c/\lab in {1/{1},2/{2},3/{3},4/{4}}{
  \node[font=\small] at (\c*0.85-0.85,2.85) {$\mathcal{H}_{\lab}$};
}
\newcommand{\srow}[3]{%
  \foreach \c in {1,...,4}{\node[cell] at (\c*0.85-0.85,#1) {};}
  \foreach \c in {#2}{\node[act] at (\c*0.85-0.85,#1) {};}
  \node[font=\small,anchor=west] at (2.85,#1) {#3};
}
\srow{2.35}{1}{$\sigma^{x}\!\otimes\! I\!\otimes\! I\!\otimes\! I$}
\srow{1.85}{3}{$I\!\otimes\! I\!\otimes\!\sigma^{z}\!\otimes\! I$}
\srow{1.35}{1,3}{$\sigma^{x}\!\otimes\! I\!\otimes\!\sigma^{z}\!\otimes\! I$}
\draw[{Latex[length=2mm]}-{Latex[length=2mm]},line width=0.9pt] (1.27,1.05) -- (1.27,-0.10);
\node[font=\scriptsize,anchor=west] at (1.55,0.475) {Eq.~\eqref{eq:substring}:};
\node[font=\scriptsize,anchor=west,align=left] at (2.65,0.475)
  {$\langle\sigma^{x}_{1}\rangle=q_{1}(0)-q_{1}(1)=-0.06$\ (row-sum diff.)\\
   $\langle\sigma^{z}_{3}\rangle=q_{3}(0)-q_{3}(1)=-0.02$\ (column-sum diff.)\\
   $\langle\sigma^{x}_{1}\sigma^{z}_{3}\rangle=q_{00}-q_{01}-q_{10}+q_{11}=0.32$\ (correlation)};
\begin{scope}[shift={(-0.3,-0.95)}]
  \fill[okgreen!12] (5.9,-0.38) rectangle (8.5,0.38);
  \fill[okgreen!12] (2.9,-0.76) rectangle (8.5,-0.38);
  \draw[cellgray,line width=0.7pt] (0,0.76) -- (8.5,0.76);
  \draw[cellgray,line width=0.7pt] (0,0.38) -- (8.5,0.38);
  \draw[cellgray!70,line width=0.3pt] (0,0.0) -- (8.5,0.0);
  \draw[cellgray!70,line width=0.3pt] (0,-0.38) -- (8.5,-0.38);
  \draw[cellgray,line width=0.7pt] (0,-0.76) -- (8.5,-0.76);
  \node[font=\scriptsize,anchor=west] at (0.1,0.57) {$q_{k_{1}k_{3}}$};
  \node[font=\scriptsize] at (3.65,0.57) {$k_{3}{=}0\,(+1)$};
  \node[font=\scriptsize] at (5.15,0.57) {$k_{3}{=}1\,(-1)$};
  \node[font=\scriptsize] at (7.2,0.57) {row sums (site $1$)};
  \node[font=\scriptsize,anchor=west] at (0.1,0.19) {$k_{1}{=}0\,(+1)$};
  \node[font=\scriptsize] at (3.65,0.19) {$q_{00}{=}0.31$};
  \node[font=\scriptsize] at (5.15,0.19) {$q_{01}{=}0.16$};
  \node[font=\scriptsize] at (7.2,0.19) {$q_{1}(0){=}0.47$};
  \node[font=\scriptsize,anchor=west] at (0.1,-0.19) {$k_{1}{=}1\,(-1)$};
  \node[font=\scriptsize] at (3.65,-0.19) {$q_{10}{=}0.18$};
  \node[font=\scriptsize] at (5.15,-0.19) {$q_{11}{=}0.35$};
  \node[font=\scriptsize] at (7.2,-0.19) {$q_{1}(1){=}0.53$};
  \node[font=\scriptsize,anchor=west] at (0.1,-0.57) {column sums (site $3$)};
  \node[font=\scriptsize] at (3.65,-0.57) {$q_{3}(0){=}0.49$};
  \node[font=\scriptsize] at (5.15,-0.57) {$q_{3}(1){=}0.51$};
  \node[font=\scriptsize] at (7.2,-0.57) {total $=1$};
\end{scope}

\node[font=\normalsize\bfseries,anchor=west] at (10.3,3.35) {(b) All slices and the max aggregation};
\begin{scope}[shift={(11.1,-1.0)}]
  \node[font=\scriptsize,anchor=east] at (-0.1,3.85) {$b\,\backslash\,S$};
  \foreach \c/\lab in {1/{\{1,2\}},2/{\{1,3\}},3/{\{1,4\}},4/{\{2,3\}},5/{\{2,4\}},6/{\{3,4\}}}{
    \node[font=\scriptsize] at (\c*0.78-0.39,3.85) {$\lab$};
  }
  \foreach \r/\lab in {1/{xx},2/{xy},3/{xz},4/{yx},5/{yy},6/{yz},7/{zx},8/{zy},9/{zz}}{
    \node[font=\scriptsize,anchor=east] at (-0.1,3.8-\r*0.4) {$\lab$};
  }
  \foreach \r in {1,...,9}{
    \foreach \c in {1,...,6}{
      \pgfmathsetmacro{\sh}{18+mod(3*\r+5*\c,7)*9}
      \draw[rounded corners=1.4pt,draw=cellgray!70,line width=0.4pt,fill=black!\sh]
        (\c*0.78-0.74,3.8-\r*0.4-0.18) rectangle (\c*0.78-0.04,3.8-\r*0.4+0.18);
    }
  }
  \draw[rounded corners=1.4pt,draw=badred,line width=1.0pt,fill=badred!85]
    (6*0.78-0.74,3.8-5*0.4-0.18) rectangle (6*0.78-0.04,3.8-5*0.4+0.18);
  \node[white,font=\scriptsize\bfseries] at (6*0.78-0.39,3.8-5*0.4) {max};
  \node[font=\small,anchor=west] at (-0.5,-0.5) {$\ell_{\tau}(u)$};
  \node[font=\scriptsize] at (0.8,-0.5) {low};
  \shade[left color=black!18,right color=black!72,draw=cellgray!70,line width=0.4pt]
    (1.1,-0.62) rectangle (2.8,-0.38);
  \node[font=\scriptsize] at (3.1,-0.5) {high};
  \draw[rounded corners=1.4pt,draw=badred,line width=1.0pt,fill=badred!85] (3.42,-0.62) rectangle (3.92,-0.38);
  \node[font=\scriptsize,anchor=east] at (4.64,-0.5) {max};
\end{scope}
\draw[-{Latex[length=2.4mm]},badred,line width=1.0pt] (15.85,0.30) -- (16.65,0.30);

\node[font=\small,anchor=west] at (16.85,2.85) {maximizing slice $\tau_{\star}=(\{3,4\},\,yy)$};
\begin{scope}[shift={(17.05,-0.98)},x=1cm,y=5.5cm]
  \draw[->,line width=0.5pt] (-0.2,0) -- (3.3,0);
  \draw[->,line width=0.5pt] (-0.2,0) -- (-0.2,0.52);
  \foreach \i/\klab/\p/\q in {0/{00}/0.42/0.31, 1/{01}/0.08/0.16, 2/{10}/0.13/0.18, 3/{11}/0.37/0.35}{
    \draw[draw=caublue!85!black,line width=0.5pt,fill=caublue!75] (\i*0.8+0.06,0) rectangle (\i*0.8+0.34,\q);
    \draw[morange,line width=1.1pt,pattern=north east lines,pattern color=morange!80]
      (\i*0.8+0.40,0) rectangle (\i*0.8+0.68,\p);
    \node[font=\small,anchor=north] at (\i*0.8+0.37,-0.008) {$\klab$};
  }
  \draw[draw=caublue!85!black,line width=0.5pt,fill=caublue!75] (0.0,0.617) rectangle (0.24,0.655);
  \node[font=\small,anchor=west] at (0.30,0.636) {$q^{\tau_{\star}}$ (data)};
  \draw[morange,line width=1.1pt,pattern=north east lines,pattern color=morange!80] (0.0,0.556) rectangle (0.24,0.594);
  \node[font=\small,anchor=west] at (0.30,0.575) {$p^{\tau_{\star}}(u)$ (model)};
  \node[font=\scriptsize,anchor=north] at (1.55,-0.075) {cell $k$};
\end{scope}
\end{tikzpicture}%
}
\vspace*{-7mm}
\caption{Structure of a slice. (a) The commuting family (slice) read out at once in the setting $\tau=(\{1,3\},(x,z))$ and the data of that setting. Data accumulate as an empirical distribution $q^{\tau}$ over cells (a cell $k$ is the tuple of readout bits of the sites in the block); the interior of the contingency table is the weight-$2$ record and the margins are its marginalization, i.e.\ the weight-$1$ data. Each string expectation is a signed sum of counts, expectations and counts being interconvertible by \eqref{eq:substring}. (b) The settings form a (basis $b$) $\times$ (block $S$) grid, of size $9\times6=54$ for $n=4$, $N=2$; the max-sliced loss picks the slice $\tau_{\star}$ (red) at which the total variation $\ell_{\tau}(u)$ (shading) between the Born prediction $p^{\tau}(u)$ and the data is worst, and its cell distribution is shown on the right. Weight-$1$ strings are diagonal in several settings, so the slices cover the measurement set with overlaps.}
\label{fig:slice}
\end{figure}
\section{Main result and sketch of its proof}\label{ssec:main}

The purpose of this section is to analyze the structure of the stationary points of the max-sliced loss $L$. The outcome is that, when the maximizer is unique, every stationary point is a saddle point.

\begin{definition}[Stationary point]\label{def:stationary}
Let $u\in\Sph$ and let $v$ be a unit tangent direction orthogonal to $u$. Write $L'(u;v)=\lim_{t\downarrow0}\bigl(L(\gamma_{v}(t))-L(u)\bigr)/t$ for the one-sided directional derivative along the great circle $\gamma_{v}(t)=\cos t\,u+\sin t\,v$ (existence is shown in the appendix). If $L'(u;v)\ge0$ in every tangent direction, $u$ is called a \emph{stationary point} of $L$.
\end{definition}

\begin{lemma}\label{lem:loss_property}
$L$ is $2$-Lipschitz on $\Sph_{d}$, and moreover $L$ is Clarke regular.
\end{lemma}

\begin{remark}
By Lemma~\ref{lem:loss_property}, $L$ is nonsmooth but globally Lipschitz, so its Clarke subdifferential is nonempty at every point \citep{clarke1990optimization}. Lipschitz continuity alone does not imply Clarke regularity (the function $-\lvert x\rvert$ is a counterexample, see Appendix~\ref{app:lipschitz}). The regularity of $L$ comes from its structure instead: $L$ is the pointwise maximum of finitely many smooth functions, and such a maximum is Clarke regular \citep[Prop.~2.3.12]{clarke1990optimization}. The stationary points of Definition~\ref{def:stationary} therefore coincide with Clarke stationary points, the standard notion in nonsmooth analysis.
\end{remark}

\begin{definition}[Support condition]\label{def:support}
Let $L(u)>0$ and let the maximizing slice $\tau_{\star}=\tau(u)$ be unique. Split the cells by the sign of the residual $\delta_{k}=\delta^{\tau_{\star}}_{k}(u)$ into $\mathcal{P}=\{k:\delta_{k}>0\}$, $\mathcal{Z}=\{k:\delta_{k}=0\}$ and $\mathcal{D}=\{k:\delta_{k}<0\}$ (mass conservation \eqref{eq:mass} implies $\mathcal{P},\mathcal{D}\neq\emptyset$). The \emph{support condition} is that every deficient cell be disjoint from the support of $u$, i.e.\ that
\begin{equation}\label{eq:support}
\Pi^{\tau_{\star}}_{k}\,u=0\qquad(\forall k\in\mathcal{D}).
\end{equation}
\end{definition}

The main result is a classification of stationary points on the stratum where the active slice is unique.

\begin{theorem}[Landscape theorem]\label{thm:main-landscape}
Let $L(u)>0$ and let the maximizing slice be unique. Then $u$ is a stationary point if and only if the support condition holds. Moreover, when the support condition holds, $u$ is not a local minimum: for any deficient cell $k_{0}\in\mathcal{D}$ and any unit vector $v\in\ran\Pi^{\tau_{\star}}_{k_{0}}$, along the great circle $\gamma_{v}$ and for all sufficiently small $t>0$,
\[
L(\gamma_{v}(t))\;=\;L(u)\;-\;2P_{\mathcal{P}}(u)\,\sin^{2}t,
\qquad
P_{\mathcal{P}}(u)=\textstyle{\sum_{k\in\mathcal{P}}}p^{\tau_{\star}}_{k}(u)\;>\;0 .
\]
\end{theorem}

\begin{corollary}\label{cor:main-ties}
Every spurious local minimum of $L$ (a local minimum with positive value) lies in the \emph{tie set}, the set of points at which several slices share the maximum. Moreover, under Assumption~\ref{as:meas} and with exact data $q^{\tau}=p^{\tau}(u_{\star})$, for almost every truth $u_{\star}$ the global minimum of $L$ is the truth up to a global phase.
\end{corollary}

\subsection{Sketch of the proof of Theorem~\ref{thm:main-landscape}}

The complete proof is in Appendix~\ref{sec:landscape}; the mechanism splits into four steps.

\emph{Step 1: localization.} Let $\Delta=\ell_{\tau_{\star}}(u)-\max_{\tau\neq\tau_{\star}}\ell_{\tau}(u)>0$ be the gap between the maximizing slice $\tau_{\star}$ and the runner-up. Since every $\ell_{\tau}$ is $2$-Lipschitz (Lemma~\ref{lem:loss_property}), none of them moves by more than $\Delta/2$ inside the ball of radius $\Delta/4$, so $\tau_{\star}$ remains the unique maximizer there. Hence $L=\ell_{\tau_{\star}}$ in that neighbourhood, and stationarity, local minimality and one-sided directional derivatives of $L$ all agree with those of the single slice $\ell_{\tau_{\star}}$.

\emph{Step 2: first order mass flows.} Putting $dp^{\tau_{\star}}_{k}(v)=2\,\mathrm{Re}\langle\Pi^{\tau_{\star}}_{k}u,v\rangle$, the $v$-directional derivative is
\[
\ell_{\tau_{\star}}'(u;v)=\sum_{k\in\Pc}dp^{\tau_{\star}}_{k}(v)-\sum_{k\in\Dc}dp^{\tau_{\star}}_{k}(v)+\sum_{k\in\Zc}\lvert dp^{\tau_{\star}}_{k}(v)\rvert,
\]
where only the absolute values on $\Zc$ carry the nonsmoothness. The first-order mass transfer is subject to two restrictions: (i) the resolution $\sum_{k}\Pi^{\tau_{\star}}_{k}=I$ and the tangency condition give $\sum_{k}dp^{\tau_{\star}}_{k}(v)=0$, so only zero-sum flows can be realized at first order; (ii) since $p^{\tau_{\star}}_{k}(u)=\lVert\Pi^{\tau_{\star}}_{k}u\rVert^{2}$ is quadratic, a cell with $p^{\tau_{\star}}_{k}(u)=0$ has $\Pi^{\tau_{\star}}_{k}u=0$ and hence $dp^{\tau_{\star}}_{k}(v)=0$ for every $v$, so no first-order inflow into an empty cell is possible. Conversely, every flow satisfying these two restrictions is realizable: for any zero-sum vector $(\alpha_{k})$ on the support $\{k:p^{\tau_{\star}}_{k}(u)>0\}$, the direction $v=\sum_{k}\frac{\alpha_{k}}{2p^{\tau_{\star}}_{k}(u)}\Pi^{\tau_{\star}}_{k}u$ realizes $dp^{\tau_{\star}}_{k}(v)=\alpha_{k}$. In short, the flows available at first order are exactly the zero-sum flows on the support.

\emph{Step 3: when the support condition fails.} Suppose a deficient cell $k_{-}\in\Dc$ carries mass. Pick a surplus cell $k_{+}\in\Pc$ (mass conservation gives $\Pc\neq\emptyset$, and on $\Pc$ we have $p^{\tau_{\star}}_{k}>q^{\tau_{\star}}_{k}\ge0$, hence $p^{\tau_{\star}}_{k}>0$). Both $k_{\pm}$ lie in the support, so by the construction of Step 2 the flow $\alpha_{k_{+}}=-\varepsilon$, $\alpha_{k_{-}}=+\varepsilon$ and $0$ elsewhere is realizable. For this $v$ every other cell has $dp^{\tau_{\star}}_{j}(v)=0$, so only two terms survive in the expression above and
\[
\ell_{\tau_{\star}}'(u;v)=dp^{\tau_{\star}}_{k_{+}}(v)-dp^{\tau_{\star}}_{k_{-}}(v)=(-\varepsilon)-(+\varepsilon)=-2\varepsilon<0 .
\]
The $\varepsilon$ removed from the surplus cell decreases $\delta^{\tau_{\star}}_{k_{+}}$ and the $\varepsilon$ poured into the deficient cell decreases $\lvert\delta^{\tau_{\star}}_{k_{-}}\rvert$. The descent comes from both sides. Hence $u$ is neither stationary nor a local minimum.

\emph{Step 4: when the support condition holds.} Now all cells of $\Dc$ are empty, so their $dp^{\tau_{\star}}_{k}$ vanish, and the same holds for $\Zc\cap\{p^{\tau_{\star}}_{k}=0\}$. Rewriting the first-order mass conservation over the surviving cells gives $\sum_{k\in\Pc}dp^{\tau_{\star}}_{k}(v)=-\sum_{k\in\Zc,\,p^{\tau_{\star}}_{k}>0}dp^{\tau_{\star}}_{k}(v)$, and substituting,
\[
\ell_{\tau_{\star}}'(u;v)=\sum_{k\in\Zc,\,p^{\tau_{\star}}_{k}>0}\bigl(\lvert dp^{\tau_{\star}}_{k}(v)\rvert-dp^{\tau_{\star}}_{k}(v)\bigr)\ \ge\ 0
\]
for every $v$, i.e.\ $u$ is stationary: lowering the loss would require pouring mass into a deficient cell, but those are empty and inflow into an empty cell does not occur at first order. At second order, however, the loss can be lowered. Take an empty deficient cell $k_{0}\in\Dc$ and a unit vector $v\in\ran\Pi^{\tau_{\star}}_{k_{0}}$. The support condition gives $\Pi^{\tau_{\star}}_{k_{0}}u=0$, hence $\langle u,v\rangle=0$ and $\Pi^{\tau_{\star}}_{k}v=0$ for $k\neq k_{0}$. By this orthogonality the projectors act diagonally along the great circle $u_{t}=\cos t\,u+\sin t\,v$, so with $h=\sin^{2}t$,
\[
p^{\tau_{\star}}_{k}(u_{t})=(1-h)\,p^{\tau_{\star}}_{k}(u)\ \ (k\neq k_{0}),
\qquad
p^{\tau_{\star}}_{k_{0}}(u_{t})=h .
\]
It remains to count the change of the absolute residuals cell by cell. In the range $h\le h_{0}:=\min\{\min_{k\in\Pc}\delta^{\tau_{\star}}_{k}/p^{\tau_{\star}}_{k},\,q^{\tau_{\star}}_{k_{0}}\}$ no residual changes sign, so cells of $\Pc$ decrease by $h\,p^{\tau_{\star}}_{k}$, cells of $\Zc\cap\{p^{\tau_{\star}}_{k}>0\}$ increase by $h\,p^{\tau_{\star}}_{k}$, the cell $k_{0}$ decreases by $h$, and the rest are unchanged. Using that the whole mass of $p$ sits on $\Pc$ and $\Zc\cap\{p^{\tau_{\star}}_{k}>0\}$, the total is
\[
\ell_{\tau_{\star}}(u_{t})-\ell_{\tau_{\star}}(u)=-h\,P_{\Pc}+h\,P_{\Zc_{+}}-h=-2h\,P_{\Pc},
\]
and $P_{\Pc}\ge\ell_{\tau_{\star}}(u)/2>0$ guarantees a strict descent. By the localization of Step 1 this identity holds for $L$ as well for all sufficiently small $t$.

\begin{figure}[t]
\centering
\resizebox{1\linewidth}{!}{%
\begin{tikzpicture}[
  x=1cm,y=1cm,
  qbar/.style={draw=caublue!85!black,line width=0.5pt,fill=caublue!70},
  pbar/.style={morange,line width=1.1pt,pattern=north east lines,pattern color=morange!85},
  flow/.style={-{Latex[length=2.2mm]},line width=1.0pt},
]
\newcommand{\moved}[3]{%
  \fill[pattern=north east lines,pattern color=cellgray] (#1,#2) rectangle (#1+0.30,#3);
  \draw[cellgray,line width=1.0pt] (#1,#2) rectangle (#1+0.30,#3);}

\node[font=\normalsize\bfseries,anchor=west] at (-0.4,3.30) {(a) support condition violated};
\begin{scope}[shift={(0,0)},x=1.02cm,y=3.9cm]
  \draw[->,line width=0.5pt] (-0.25,0) -- (4.35,0);
  \draw[->,line width=0.5pt] (-0.25,0) -- (-0.25,0.72);
  \foreach \i/\q/\p in {0/0.22/0.32, 1/0.18/0.26, 2/0.26/0.26, 3/0.34/0.16}{
    \draw[qbar] (\i+0.10,0) rectangle (\i+0.40,\q);
    \draw[pbar] (\i+0.46,0) rectangle (\i+0.76,\p);}
  \foreach \i/\lab in {0/{$\mathcal{P}$},1/{$\mathcal{P}$},2/{$\mathcal{Z}$},3/{$\mathcal{D}$}}{
    \node[font=\small,anchor=north] at (\i+0.43,-0.012) {\lab};}
  \draw[flow,badred] (0.61,0.345) to[out=55,in=125] (3.61,0.195);
  \node[badred,font=\small,anchor=south] at (2.1,0.575) {move $\varepsilon$ to a deficient cell};
  \node[badred,font=\small,anchor=south] at (2.1,0.655) {$\ell_{\tau}'(u;v)=-2\varepsilon<0$};
\end{scope}

\node[font=\normalsize\bfseries,anchor=west] at (6.6,3.30) {(b-1) first order: no motion};
\begin{scope}[shift={(7.0,0)},x=1.02cm,y=3.9cm]
  \draw[->,line width=0.5pt] (-0.25,0) -- (4.35,0);
  \draw[->,line width=0.5pt] (-0.25,0) -- (-0.25,0.72);
  \foreach \i/\q/\p in {0/0.22/0.40, 1/0.18/0.34, 2/0.26/0.26, 3/0.34/0.00}{
    \draw[qbar] (\i+0.10,0) rectangle (\i+0.40,\q);
    \ifdim \p pt > 0pt \draw[pbar] (\i+0.46,0) rectangle (\i+0.76,\p); \fi}
  \foreach \i/\lab in {0/{$\mathcal{P}$},1/{$\mathcal{P}$},2/{$\mathcal{Z}$},3/{$\mathcal{D}$}}{
    \node[font=\small,anchor=north] at (\i+0.43,-0.012) {\lab};}
  \node[badred,font=\small,anchor=north] at (3.61,-0.075) {$p_{k_{0}}=0$};
  \draw[flow,okgreen] (0.61,0.425) to[out=55,in=125] (1.61,0.365);
  \draw[flow,okgreen] (1.61,0.365) to[out=55,in=125] (2.61,0.285);
  \draw[flow,badred,dashed] (2.61,0.285) to[out=45,in=135] (3.61,0.055);
  \node[badred,font=\small] at (3.40,0.395) {no inflow};
  \node[okgreen,font=\small,anchor=south] at (1.9,0.575) {transfers inside the support};
  \node[okgreen,font=\small,anchor=south] at (1.6,0.655) {$\ell_{\tau}'(u;v)\ge0$};
\end{scope}

\node[font=\normalsize\bfseries,anchor=west] at (13.2,3.30) {(b-2) second order: descent};
\begin{scope}[shift={(13.6,0)},x=1.02cm,y=3.9cm]
  \draw[->,line width=0.5pt] (-0.25,0) -- (4.35,0);
  \draw[->,line width=0.5pt] (-0.25,0) -- (-0.25,0.72);
  \foreach \i/\q/\p in {0/0.22/0.3520, 1/0.18/0.2992, 2/0.26/0.2288, 3/0.34/0.00}{
    \draw[qbar] (\i+0.10,0) rectangle (\i+0.40,\q);
    \ifdim \p pt > 0pt \draw[pbar] (\i+0.46,0) rectangle (\i+0.76,\p); \fi}
  \moved{0.46}{0.3520}{0.40}
  \moved{1.46}{0.2992}{0.34}
  \moved{2.46}{0.2288}{0.26}
  \moved{3.46}{0}{0.12}
  \foreach \i/\lab in {0/{$\mathcal{P}$},1/{$\mathcal{P}$},2/{$\mathcal{Z}$},3/{$\mathcal{D}$}}{
    \node[font=\small,anchor=north] at (\i+0.43,-0.012) {\lab};}
  \node[font=\small,anchor=south,align=center] at (1.9,0.575)
    {grey blocks move (total $h$)};
  \node[caublue!60!black,font=\small,anchor=south] at (1.9,0.655)
    {$\ell_{\tau}(u_{t})=\ell_{\tau}(u)-2P_{\mathcal{P}}h$};
\end{scope}
\draw[flow,caublue!60!black] (15.60,1.70) to[out=-8,in=95] (17.28,0.55);
\node[caublue!60!black,font=\small,anchor=south west] at (15.58,1.74) {inflow $h=\sin^{2}t$};
\end{tikzpicture}%
}
\vspace*{-8mm}
\caption{The support condition and the order of the mass flow. Blue is the data $q^{\tau}$ and orange hatching the model $p^{\tau}(u)$. Cells are split by the sign of the residual $\delta_{k}=p_{k}-q_{k}$ into surplus $\mathcal{P}$, zero $\mathcal{Z}$ and deficient $\mathcal{D}$, and what can be moved at first order is exactly a zero-sum flow on the support ($p_{k}>0$) of the model. (a) When the support condition fails, a deficient cell carries mass, so the flow moving $\varepsilon$ from a surplus cell to it is realizable and the loss decreases strictly: the point is not stationary. (b-1) Under the support condition the deficient cell is empty, no first-order inflow into it is possible, and only transfers inside the support remain, none of which decreases the loss in any direction: $u$ is stationary. (b-2) Along the great circle $u_{t}=\cos t\,u+\sin t\,v$ with $v\in\ran\Pi^{\tau}_{k_{0}}$, however, the inflow into the empty cell, $p_{k_{0}}(u_{t})=\sin^{2}t=:h$, is achieved at second order in $t$ while every support cell is scaled by $(1-h)$; the grey hatched blocks are the moving mass. As long as no residual changes sign this strictly decreases the loss, so $u$ is not a local minimum.}
\label{fig:support}
\end{figure}
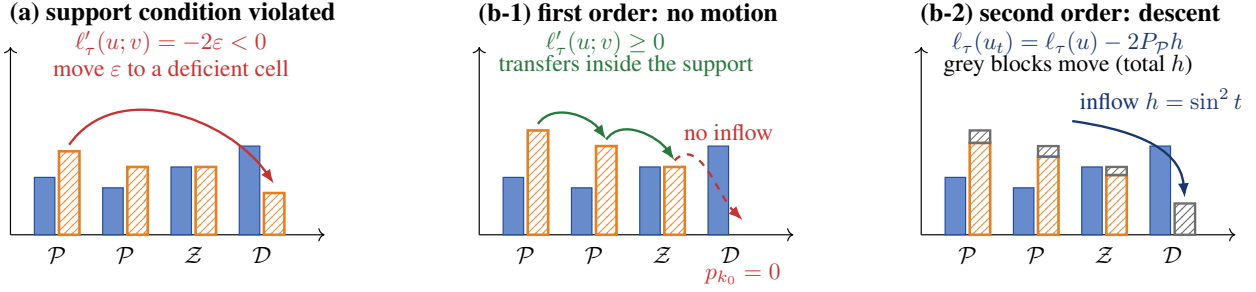
\section{Algorithms and experiments}\label{sec:experiments}

\paragraph{Data generation.}
All experiments use $n=4$ ($N=2$, pair blocks, $d=16$). As a target that empirically tends to create traps we start from a state with equal amplitudes on all computational basis vectors, differing only in sign. Writing $e_{a}$ ($a\in\{0,1\}^{n}$) for the computational basis and $f(a)=\sum_{(i,j)\in E}a_{i}a_{j}\ (\mathrm{mod}\ 2)$ for the quadratic form determined by an edge set $E$, put
\begin{equation}\label{eq:graphstate}
\psi_{0}\;=\;\frac{1}{\sqrt d}\sum_{a\in\{0,1\}^{n}}(-1)^{f(a)}\,e_{a}.
\end{equation}
In quantum information this family is called a graph state, and taking $E$ to be a cycle on $n$ vertices gives the ring cluster state. For $n=4$ and $E=\{(1,2),(2,3),(3,4),(4,1)\}$ only $a\in\{0011,0110,1001,1100\}$ carry a negative sign. This family is highly symmetric and loses unique determination under weight-restricted measurements. Into it we mix a phased $W$-type state $W=(e_{0001}+\mathrm{i}\,e_{0010}-e_{0100}-\mathrm{i}\,e_{1000})/2$ orthogonal to $\psi_{0}$, forming the reference state
\[
\psi^{\star}\ \propto\ 4\,\psi_{0}+W,\qquad \lvert\psi_{0}^{\dagger}\psi^{\star}\rvert^{2}=\tfrac{16}{17},
\]
whose $W$ component decisively breaks the symmetry of $\psi_{0}$. In addition, for each trial we draw an independent isotropic complex Gaussian unit direction $g$ orthogonal to $\psi^{\star}$ and take as the target
\begin{equation}\label{eq:perturb}
\psi_{\varepsilon}\;=\;\frac{\psi^{\star}+\varepsilon\,g}{\lVert\psi^{\star}+\varepsilon\,g\rVert},\qquad \varepsilon=0.05
\end{equation}
(with $F(\psi_{\varepsilon},\psi^{\star})=1/(1+\varepsilon^{2})\approx0.998$). The perturbed target lands at a generic point and falls into the region of unique determination on the pure-state manifold in the sense of Corollary~\ref{cor:main-ties}, whereas the $\ell^{2}$ landscape is not benign. The target is uniquely determined and yet least squares gets trapped, exactly the situation envisaged by Theorem~\ref{thm:main-landscape}. Physically it corresponds to a state prepared on hardware that deviates slightly from the ideal symmetric state. Fidelity is measured by $F(\varphi,\psi)=\lvert\psi^{\dagger}\varphi\rvert^{2}$, which is invariant under the global phase.

\paragraph{Algorithm.}
The slice family has $K=54$ members ($6$ pairs $\times$ $9$ bases). To minimize the loss $L$ we use Polyak subgradient with momentum, driven by the signed subgradient $g$ of the active slice,
\[
w\ \leftarrow\ \beta\,w-\frac{L(u)}{\lVert g\rVert^{2}}\,g,
\qquad
u\ \leftarrow\ \frac{u+w}{\lVert u+w\rVert},
\qquad \beta=0.5 .
\]
Here $g=2\sum_{k}\operatorname{sign}\bigl(\delta^{\tau_{\star}}_{k}(u)\bigr)\,\Pi^{\tau_{\star}}_{k}u$ for the maximizing slice $\tau_{\star}$, an element of the Clarke subdifferential $\partial_{C}L(u)$. Initial points are uniform on the sphere, and the baseline factored least squares (FGD) is run from the same data and the same initial point. Data are generated as in Assumption~\ref{as:meas}. Detection efficiency and drift are not simulated---they motivate the weight restriction (\S\ref{sec:intro}), and the restricted measurement is treated as ideal multinomial counting.

\paragraph{Results.}
We first compare, on exact data ($R=\infty$), four losses that differ only in how the residuals are aggregated (Table~\ref{tab:exact}). The $\ell^{2}$ sum of Pauli residuals is trapped at a spurious local minimum ($F\approx0$) in $25$ of $100$ trials, and replacing the residuals by cell distributions, or by total variation, does not remove the traps. They disappear when the sum is replaced by a max. What matters is not the choice of distance but the change of aggregation from sum to max.

\begin{table}[t]
\centering\small
\begin{tabular}{llc}
\hline
Loss (aggregation) & Definition & Success rate ($F>0.999$) \\
\hline
$\ell^{2}$ sum of Pauli residuals (FGD)
  & $\|\A(uu^{\dagger})-y\|_{2}^{2}$ & $75/100$ \\
$\ell^{2}$ sum of cell residuals
  & $\sum_{\tau}\lVert p^{\tau}(u)-q^{\tau}\rVert_{2}^{2}$ & $94/100$ \\
Sum of slice total variations
  & $\sum_{\tau}\lVert p^{\tau}(u)-q^{\tau}\rVert_{1}$ & $74/100$ \\
Max-sliced total variation \eqref{eq:loss}
  & $\max_{\tau}\lVert p^{\tau}(u)-q^{\tau}\rVert_{1}$ & $\mathbf{100/100}$ \\
\hline
\end{tabular}
\caption{Identical noiseless data and identical initial points. The first three rows all sum over $\tau$, only the last takes a max. The max-sliced loss reaches $F=1.0000$ in every trial.}
\label{tab:exact}
\end{table}
The finite-shot comparison is given in Table~\ref{tab:noise}. The max-sliced loss has no traps at $R=10^{4},10^{5}$ (minimum fidelity $0.994$ and $0.999$), but because of the statistical floor its final accuracy is slightly below the successful runs of $\ell^{2}$. A \emph{two-stage} scheme---descend with the max-sliced loss, then polish with $\ell^{2}$---combines the absence of traps with the accuracy and is best at every shot budget. At $R=10^{3}$ the max-sliced loss alone does not reach the threshold $F>0.99$ (median $0.972$), and a few of the $40$ runs stall in a near-tie band.

\begin{table}[t]
\centering\small
\begin{tabular}{lcccccc}
\hline
& \multicolumn{2}{c}{$\ell^{2}$ (FGD)} & \multicolumn{2}{c}{max-sliced} & \multicolumn{2}{c}{two-stage} \\
$R$ & success & median $F$ & success & median $F$ & success & median $F$ \\
\hline
$10^{5}$ & $0.75$ & $0.9999$ & $1.00$ & $0.9997$ & $\mathbf{1.00}$ & $\mathbf{0.9999}$ \\
$10^{4}$ & $0.72$ & $0.9990$ & $1.00$ & $0.9971$ & $\mathbf{1.00}$ & $\mathbf{0.9993}$ \\
$10^{3}$ & $0.35$ & $0.9717$ & $0.00$ & $0.9722$ & $\mathbf{0.65}$ & $\mathbf{0.9937}$ \\
\hline
\end{tabular}
\caption{Finite-shot comparison ($40$ trials each, $K=54$; success means $F>0.99$). Two-stage $=$ $\ell^{2}$ polishing initialized at the endpoint of the max-sliced descent.}
\label{tab:noise}
\end{table}
\section{Discussion}\label{sec:discussion}
Theorem~\ref{thm:main-landscape} classifies the stationary points on the stratum where the maximizing slice is unique, and Corollary~\ref{cor:main-ties} confines spurious local minima to the tie set, where several slices share the maximum. In practice an exact tie is never attained because of numerical error. A Clarke stationary point on the tie set is therefore seen in computation not as a point where the iterates come to rest but as an oscillation between the slices that compete for the maximum, and momentum largely suppresses this oscillation: in the setting of \S\ref{sec:experiments} the success rate improves from $72/100$ without momentum to $100/100$ with it. A rigorous analysis of the stationary points on the tie set is left for future work.

\subsection*{Acknowledgments}

This work was supported by ERATO ``Super Quantum Entanglement'' (Grant No. JPMJER2402) from JST.

\subsection*{AI use statement}

In this work, we used generative AI tools for implementing the numerical experiments, for translating and copy-editing the manuscript, and for producing the figure sources. We have not used generative AI tools for research ideation, for formulating the theoretical claims, or for constructing their proofs; the definitions, the statements and the proofs in \S\ref{ssec:main} and Appendix~\ref{sec:landscape} were written and verified by the author. We have reviewed all AI-assisted work: the experiment code was inspected and re-run by the author, every number reported in \S\ref{sec:experiments} was reproduced from that code, and all references were checked against their primary sources. We take responsibility for the final content of this work, including text, claims or artifacts produced with the aid of generative AI.

\appendix
\section{Proof of the landscape theorem}\label{sec:landscape}

In this appendix we restate Lemma~\ref{lem:loss_property}, Theorem~\ref{thm:main-landscape} and Corollary~\ref{cor:main-ties} of the main text under their own numbers and prove them. The organization follows the order of the proof sketch in \S\ref{ssec:main}: after the preliminaries and the proof of Lemma~\ref{lem:loss_property}, Step 1 (localization) is Lemma~\ref{lem:localize}, Step 2 (mass flows realizable at first order) is Lemma~\ref{lem:dd}, Step 3 (the support condition fails) is Lemma~\ref{lem:violate} and Step 4 (the support condition holds) is Lemma~\ref{lem:supp}; combining them proves Theorem~\ref{thm:main-landscape} and Corollary~\ref{cor:main-ties}.

\subsection{Preliminaries}\label{app:prelim}

Throughout, $\Sph=\{u\in\C^{d}:\lVert u\rVert=1\}$, and the \emph{real tangent space} at $u\in\Sph$ is
\[
T_{u}=\bigl\{v\in\C^{d}:\ \mathrm{Re}\langle u,v\rangle=0\bigr\},
\]
the set of directions orthogonal to $u$ in the real part of the complex inner product, a real linear space of dimension $2d-1$. A ``unit tangent direction orthogonal to $u$'' in Definition~\ref{def:stationary} is a unit vector of $T_{u}$, and for such a $v$ the great circle $\gamma_{v}(t)=\cos t\,u+\sin t\,v$ lies on $\Sph$ because $\lVert\gamma_{v}(t)\rVert^{2}=\cos^{2}t+\sin^{2}t+2\sin t\cos t\,\mathrm{Re}\langle u,v\rangle=1$. The one-sided directional derivative $f'(u;v)=\lim_{t\downarrow0}\bigl[f(\gamma_{v}(t))-f(u)\bigr]/t$ of $f:\Sph\to\R$ and the notion of a stationary point are as in Definition~\ref{def:stationary}, and we use the same words for each slice loss $\ell_{\tau}$ as well as for $L$.

Fix a slice $\tau$, write its projectors as $\{\Pi^{\tau}_{k}\}_{k}$ and its cell probabilities as $p^{\tau}_{k}(u)=u^{\dagger}\Pi^{\tau}_{k}u$. Since $\Pi^{\tau}_{k}$ is Hermitian and idempotent, $p^{\tau}_{k}(u)=\lVert\Pi^{\tau}_{k}u\rVert^{2}$, and the resolution of the identity \eqref{eq:resolution} gives the Parseval-type identity $\sum_{k}\lVert\Pi^{\tau}_{k}x\rVert^{2}=\lVert x\rVert^{2}$ for every $x\in\C^{d}$. In particular $\sum_{k}p^{\tau}_{k}(u)=\lVert u\rVert^{2}=1$, while $q^{\tau}$ consists of relative frequencies so that $\sum_{k}q^{\tau}_{k}=1$; subtracting gives the mass conservation \eqref{eq:mass}. The \emph{support} of $u$ is
\[
\mathrm{supp}(u)=\{k:\ p^{\tau}_{k}(u)>0\},
\]
and the first-order change of the cell probabilities induced by $v\in T_{u}$ is written
\[
dp^{\tau}_{k}(v)=2\,\mathrm{Re}\langle\Pi^{\tau}_{k}u,\,v\rangle .
\]
Each $dp^{\tau}_{k}$ is a real linear functional on $T_{u}$; that it really is the first-order change is verified in Lemma~\ref{lem:dd}. The partition of the cells by the sign of the residual $\delta^{\tau}_{k}(u)=p^{\tau}_{k}(u)-q^{\tau}_{k}$ into $\Pc=\{k:\delta^{\tau}_{k}(u)>0\}$, $\Zc=\{k:\delta^{\tau}_{k}(u)=0\}$ and $\Dc=\{k:\delta^{\tau}_{k}(u)<0\}$, and the support condition $\Pi^{\tau}_{k}u=0$ for all $k\in\Dc$, are as in Definition~\ref{def:support}, always referring to the residuals of the slice in question. Since $p^{\tau}_{k}(u)=\lVert\Pi^{\tau}_{k}u\rVert^{2}$, the support condition is equivalent to $p^{\tau}_{k}(u)=0$ for every cell of $\Dc$.

\subsection{Proof of Lemma~\ref{lem:loss_property}}\label{app:lipschitz}

The loss $L$ is not differentiable, so stationarity cannot be written with a gradient; the tool we use instead is Clarke's nonsmooth analysis \citep{clarke1990optimization}. For a locally Lipschitz $F$ and a direction $v$, the \emph{Clarke generalized directional derivative} and the (ordinary) \emph{one-sided directional derivative} are
\[
F^{\circ}(u;v)=\limsup_{w\to u,\ t\downarrow0}\frac{F(w+tv)-F(w)}{t},
\qquad
F'(u;v)=\lim_{t\downarrow0}\frac{F(u+tv)-F(u)}{t}.
\]
The former is an upper limit in which the base point $w$ also moves; Lipschitz continuity makes it finite and convex in $v$, and it supports the \emph{Clarke subdifferential} $\partial_{C}F(u)=\{\xi:\langle\xi,v\rangle\le F^{\circ}(u;v)\ \forall v\}$, which is nonempty, compact and convex with $F^{\circ}(u;v)=\max_{\xi\in\partial_{C}F(u)}\langle\xi,v\rangle$. The latter is the naive derivative at a fixed base point and need not exist. If $F'(u;v)$ exists in every direction at $u$ and $F'(u;v)=F^{\circ}(u;v)$ for all $v$, then $F$ is said to be \emph{Clarke regular} at $u$. Regularity is what makes the two notions of stationarity agree: if $F$ is regular then
\[
0\in\partial_{C}F(u)\iff F'(u;v)\ge0\ \ (\forall v),
\]
so the stationarity of Definition~\ref{def:stationary}, defined through one-sided directional derivatives, can be read directly as the standard stationarity $0\in\partial_{C}F(u)$ of nonsmooth optimization. Without regularity the equivalence fails: for $F(x)=-\lvert x\rvert$ at $x=0$ we have $F'(0;\pm1)=-1<0$, so the point is not stationary in the directional sense, yet $F^{\circ}(0;v)=\lvert v\rvert$ gives $\partial_{C}F(0)=[-1,1]\ni0$, so it is Clarke stationary.

\begin{customlem}{\ref*{lem:loss_property}}
$L$ is $2$-Lipschitz on $\Sph_{d}$. Moreover $L$ is Clarke regular.
\end{customlem}

\begin{proof}
We first prove the Lipschitz property. Since $L$ is the maximum of $\{\ell_{\tau}\}_{\tau}$ we have $\lvert L(u)-L(w)\rvert\le\max_{\tau}\lvert\ell_{\tau}(u)-\ell_{\tau}(w)\rvert$, so it suffices to show that each $\ell_{\tau}$ is $2$-Lipschitz; this fact is used again in Lemma~\ref{lem:localize}. Fix $\tau$. The triangle inequality for norms gives $\lvert\ell_{\tau}(u)-\ell_{\tau}(w)\rvert\le\lVert p^{\tau}(u)-p^{\tau}(w)\rVert_{1}$, so we bound the right-hand side. As $p^{\tau}_{k}(u)=\lVert\Pi^{\tau}_{k}u\rVert^{2}$, factoring the difference of squares gives
\[
\bigl\lvert p^{\tau}_{k}(u)-p^{\tau}_{k}(w)\bigr\rvert
=\bigl\lvert\,\lVert\Pi^{\tau}_{k}u\rVert-\lVert\Pi^{\tau}_{k}w\rVert\,\bigr\rvert
\cdot\bigl(\lVert\Pi^{\tau}_{k}u\rVert+\lVert\Pi^{\tau}_{k}w\rVert\bigr).
\]
Summing over $k$ and applying the Cauchy--Schwarz inequality to the two factors,
\[
\lVert p^{\tau}(u)-p^{\tau}(w)\rVert_{1}
\le
\Bigl(\sum_{k}\bigl\lvert\,\lVert\Pi^{\tau}_{k}u\rVert-\lVert\Pi^{\tau}_{k}w\rVert\,\bigr\rvert^{2}\Bigr)^{1/2}
\Bigl(\sum_{k}\bigl(\lVert\Pi^{\tau}_{k}u\rVert+\lVert\Pi^{\tau}_{k}w\rVert\bigr)^{2}\Bigr)^{1/2}.
\]
The first factor is at most $\lVert u-w\rVert$ by the reverse triangle inequality $\bigl\lvert\lVert\Pi^{\tau}_{k}u\rVert-\lVert\Pi^{\tau}_{k}w\rVert\bigr\rvert\le\lVert\Pi^{\tau}_{k}(u-w)\rVert$ together with the Parseval-type identity of \S\ref{app:prelim}, and the second is at most $2$ because $(a+b)^{2}\le2(a^{2}+b^{2})$ and the same identity give $\sum_{k}(\lVert\Pi^{\tau}_{k}u\rVert+\lVert\Pi^{\tau}_{k}w\rVert)^{2}\le2(\lVert u\rVert^{2}+\lVert w\rVert^{2})=4$. Together, $\lVert p^{\tau}(u)-p^{\tau}(w)\rVert_{1}\le2\lVert u-w\rVert$. The number of cells is absorbed by the Parseval-type identity, so the constant does not depend on it, and a pointwise maximum inherits the largest of the individual Lipschitz constants, so the constant for $L$ does not depend on the number of slices either. This estimate never uses what $q^{\tau}$ is, so it holds even when the empirical distribution is not generated by any state; and since the geodesic distance on $\Sph_{d}$ dominates the chordal distance $\lVert u-w\rVert$, $L$ is $2$-Lipschitz for the geodesic distance as well.

We next prove Clarke regularity. Rewrite the absolute values as a maximum over sign patterns: let $m=2^{N}$ be the number of cells of a slice and, for a sign vector $s\in\{\pm1\}^{m}$, put
\[
f_{\tau,s}(u)\;=\;\sum_{k}s_{k}\bigl(p^{\tau}_{k}(u)-q^{\tau}_{k}\bigr).
\]
Then $\sum_{k}\lvert\delta^{\tau}_{k}(u)\rvert=\max_{s}f_{\tau,s}(u)$, so
\begin{equation}\label{eq:maxrep}
L(u)\;=\;\max_{\tau}\max_{s\in\{\pm1\}^{m}}f_{\tau,s}(u)
\end{equation}
exhibits $L$ as a maximum of finitely many ($K\,2^{m}$, where $K$ is the number of slices) functions. Each $f_{\tau,s}$ is, upon identifying $u$ with the real $2d$-dimensional vector $(\operatorname{Re}u,\operatorname{Im}u)$, a real quadratic form with constant coefficients, hence a $C^{\infty}$ function with gradient $\nabla f_{\tau,s}(u)=2\sum_{k}s_{k}\Pi^{\tau}_{k}u$. A $C^{\infty}$ function is Clarke regular, so the proposition on pointwise maxima \citep[Prop.~2.3.12]{clarke1990optimization}---if all active components are regular then the maximum is regular and its subdifferential is the convex hull of the subdifferentials of the active components---applies to \eqref{eq:maxrep}: $L$ is Clarke regular at every point and, with the active set $A(u)=\{(\tau,s):f_{\tau,s}(u)=L(u)\}$,
\[
\partial_{C}L(u)\;=\;\operatorname{co}\bigl\{\nabla f_{\tau,s}(u):(\tau,s)\in A(u)\bigr\}.
\]
In particular the Clarke generalized directional derivative coincides with the one-sided directional derivative and $L'(u;v)=\max_{(\tau,s)\in A(u)}\langle\nabla f_{\tau,s}(u),v\rangle$. Neither $\ell_{\tau}$ nor $L$ is $C^{1}$---both bend at cells where a residual changes sign---but regularity coexists with this nonsmoothness: all of it comes from the maximum in \eqref{eq:maxrep}, and the functions underneath the maximum are smooth.
\end{proof}

\subsection{Proof of the main theorem}

\subsubsection*{Step 1: localization}\label{app:localize}

\begin{lemma}[Localization]\label{lem:localize}
Let $u\in\Sph$ satisfy $L(u)>0$ and let the maximizing slice $\tau_{\star}$ be unique, and put $\Delta=\ell_{\tau_{\star}}(u)-\max_{\tau\neq\tau_{\star}}\ell_{\tau}(u)>0$. Then $\tau_{\star}$ is the unique maximizing slice at every $w\in\Sph$ with $\lVert w-u\rVert<\Delta/4$, and $L(w)=\ell_{\tau_{\star}}(w)$ there. Consequently, at $u$ the one-sided directional derivatives of $L$ and of $\ell_{\tau_{\star}}$ agree for every unit vector $v\in T_{u}$ (one exists if and only if the other does), and $u$ is a stationary point of $L$ if and only if it is one of $\ell_{\tau_{\star}}$, and a local minimum of $L$ if and only if it is one of $\ell_{\tau_{\star}}$.
\end{lemma}

\begin{proof}
As shown in the proof of Lemma~\ref{lem:loss_property}, every $\ell_{\tau}$ is $2$-Lipschitz, so $\lVert w-u\rVert<\Delta/4$ implies $\lvert\ell_{\tau}(w)-\ell_{\tau}(u)\rvert<\Delta/2$ for every $\tau$. Hence
\[
\ell_{\tau_{\star}}(w)\ >\ \ell_{\tau_{\star}}(u)-\tfrac{\Delta}{2}\ =\ \max_{\tau\ne\tau_{\star}}\ell_{\tau}(u)+\tfrac{\Delta}{2}\ >\ \max_{\tau\ne\tau_{\star}}\ell_{\tau}(w),
\]
so $\tau_{\star}$ is still the unique maximizing slice at $w$ and $L(w)=\ell_{\tau_{\star}}(w)$. Since $\gamma_{v}(t)\to u$ as $t\downarrow0$, for small $t>0$ the point $\gamma_{v}(t)$ lies in this neighbourhood and the difference quotients defining the one-sided directional derivatives of $L$ and $\ell_{\tau_{\star}}$ coincide. Stationarity is defined by the sign of those derivatives and therefore agrees, and local minimality compares function values in a neighbourhood of $u$, on which $L=\ell_{\tau_{\star}}$, so it agrees as well.
\end{proof}

\subsubsection*{Step 2: mass flows realizable at first order}\label{app:flow}

\begin{lemma}[Directional derivative and realizable mass flows]\label{lem:dd}
Fix $u\in\Sph$ and a slice $\tau$. First, for a unit vector $v\in T_{u}$ and each cell,
\[
p^{\tau}_{k}(\gamma_{v}(t))=p^{\tau}_{k}(u)+t\,dp^{\tau}_{k}(v)+O(t^{2}),
\]
with a remainder uniform in $k$. Second, the one-sided directional derivative $\ell_{\tau}'(u;v)$ exists and
\begin{equation}\label{eq:dd}
\ell_{\tau}'(u;v)
=\sum_{k\in\Pc}dp^{\tau}_{k}(v)-\sum_{k\in\Dc}dp^{\tau}_{k}(v)+\sum_{k\in\Zc}\bigl\lvert dp^{\tau}_{k}(v)\bigr\rvert,
\end{equation}
and $L'(u;v)$ exists as well, with $L'(u;v)=\max_{\tau:\,\ell_{\tau}(u)=L(u)}\ell_{\tau}'(u;v)$ (this is the existence announced in Definition~\ref{def:stationary}). Third, the set of mass flows realizable at first order is exactly the set of zero-sum vectors on the support:
\begin{equation}\label{eq:flowset}
\bigl\{\,(dp^{\tau}_{k}(v))_{k}\ :\ v\in T_{u}\,\bigr\}
\;=\;
\Bigl\{\,(\alpha_{k})_{k}\ :\ \sum_{k}\alpha_{k}=0,\ \ \alpha_{k}=0\ \ (\forall k\notin\mathrm{supp}(u))\,\Bigr\}.
\end{equation}
In particular the first-order mass conservation $\sum_{k}dp^{\tau}_{k}(v)=0$ holds for every $v\in T_{u}$, cells with $p^{\tau}_{k}(u)=0$ have $dp^{\tau}_{k}\equiv0$, and conversely every zero-sum vector $(\alpha_{k})$ on the support is realized as $dp^{\tau}_{k}(v)=\alpha_{k}$ for all $k$ by
\[
v=\sum_{k\in\mathrm{supp}(u)}\frac{\alpha_{k}}{2\,p^{\tau}_{k}(u)}\,\Pi^{\tau}_{k}u\ \in\ T_{u}.
\]
\end{lemma}

\begin{proof}
First. Since $\Pi^{\tau}_{k}$ is Hermitian and idempotent and $\mathrm{Re}\langle u,v\rangle=0$, along $\gamma_{v}(t)=\cos t\,u+\sin t\,v$,
\[
p^{\tau}_{k}(\gamma_{v}(t))
=\cos^{2}t\,p^{\tau}_{k}(u)+2\sin t\cos t\,\mathrm{Re}\langle\Pi^{\tau}_{k}u,v\rangle+\sin^{2}t\,\langle v,\Pi^{\tau}_{k}v\rangle
=p^{\tau}_{k}(u)+t\,dp^{\tau}_{k}(v)+O(t^{2}).
\]
The bounds $0\le p^{\tau}_{k}\le1$, $\lvert dp^{\tau}_{k}(v)\rvert\le2$ and $0\le\langle v,\Pi^{\tau}_{k}v\rangle\le1$ do not depend on $k$, so the remainder is $O(t^{2})$ uniformly in $k$.

Second. On cells with $\delta^{\tau}_{k}\ne0$, the residual $\delta^{\tau}_{k}(\gamma_{v}(t))=\delta^{\tau}_{k}+t\,dp^{\tau}_{k}(v)+O(t^{2})$ keeps the sign of $\delta^{\tau}_{k}$ for small $t>0$, so the one-sided derivative of $\lvert\delta^{\tau}_{k}\rvert$ is $\mathrm{sgn}(\delta^{\tau}_{k})\,dp^{\tau}_{k}(v)$; on cells with $\delta^{\tau}_{k}=0$ we have $\lvert\delta^{\tau}_{k}(\gamma_{v}(t))\rvert/t=\lvert dp^{\tau}_{k}(v)+O(t)\rvert\to\lvert dp^{\tau}_{k}(v)\rvert$. As there are finitely many cells, sum and limit commute and \eqref{eq:dd} follows. For $L$: a slice that is not active ($\ell_{\tau}(u)<L(u)$) does not contribute to the maximum for small $t$ by continuity, so $L(\gamma_{v}(t))=\max_{\tau\ \text{active}}\ell_{\tau}(\gamma_{v}(t))$, and since $\ell_{\tau}(\gamma_{v}(t))=L(u)+t\,\ell_{\tau}'(u;v)+o(t)$ for each active $\tau$, taking the maximum of finitely many such expressions gives $L(\gamma_{v}(t))-L(u)=t\max_{\tau\ \text{active}}\ell_{\tau}'(u;v)+o(t)$, so $L'(u;v)$ exists and equals the stated expression.

Third. For the inclusion $\subseteq$: the resolution of the identity \eqref{eq:resolution} and the tangency condition give
\[
\sum_{k}dp^{\tau}_{k}(v)=2\,\mathrm{Re}\bigl\langle\bigl(\textstyle\sum_{k}\Pi^{\tau}_{k}\bigr)u,\ v\bigr\rangle=2\,\mathrm{Re}\langle u,v\rangle=0,
\]
so the vector is zero-sum; and $p^{\tau}_{k}(u)=\lVert\Pi^{\tau}_{k}u\rVert^{2}$ means that $p^{\tau}_{k}(u)=0$ implies $\Pi^{\tau}_{k}u=0$, whence $dp^{\tau}_{k}(v)=2\,\mathrm{Re}\langle\Pi^{\tau}_{k}u,v\rangle=0$ for every $v$, i.e.\ the components outside the support vanish. For $\supseteq$: take any element $(\alpha_{k})$ of the right-hand side and let $v$ be as in the statement. By orthogonality of the projectors, $\{\Pi^{\tau}_{k}u\}_{k\in\mathrm{supp}(u)}$ are mutually orthogonal with $\lVert\Pi^{\tau}_{k}u\rVert^{2}=p^{\tau}_{k}(u)$, so
\[
\mathrm{Re}\langle u,v\rangle=\sum_{k\in\mathrm{supp}(u)}\frac{\alpha_{k}}{2p^{\tau}_{k}(u)}\,\mathrm{Re}\langle u,\Pi^{\tau}_{k}u\rangle=\sum_{k}\frac{\alpha_{k}}{2}=0,
\]
that is, $v\in T_{u}$. Furthermore $dp^{\tau}_{j}(v)=2\,\mathrm{Re}\langle\Pi^{\tau}_{j}u,v\rangle=\alpha_{j}$ for $j\in\mathrm{supp}(u)$, while $\Pi^{\tau}_{j}u=0$ gives $dp^{\tau}_{j}(v)=0=\alpha_{j}$ otherwise. Hence $(dp^{\tau}_{k}(v))_{k}=(\alpha_{k})_{k}$, and both inclusions give the equality \eqref{eq:flowset}.
\end{proof}

The equality \eqref{eq:flowset} supports the argument below in two directions. The inclusion $\subseteq$ provides the \emph{restriction} that no mass can be poured into an empty cell at first order, which is the reason for stationarity at points satisfying the support condition (Step 4). The inclusion $\supseteq$ provides the \emph{freedom} that any zero-sum flow on the support is realizable, which is used to construct the descent direction moving mass from a surplus cell to a deficient one when the support condition fails (Step 3). Note also that $dp^{\tau}_{k}$ is linear in $v$ and the right-hand side of \eqref{eq:dd} is positively homogeneous in $v$, so even if the realizing $v$ is not a unit vector, the sign of the one-sided directional derivative along $v/\lVert v\rVert$ is unchanged.

\subsubsection*{Step 3: when the support condition fails}\label{app:violate}

\begin{lemma}[A point violating the support condition is not stationary]\label{lem:violate}
Fix a slice $\tau$ and let $u\in\Sph$ satisfy $\ell_{\tau}(u)>0$. If the support condition fails, i.e.\ if some deficient cell $k_{-}\in\Dc$ has $p^{\tau}_{k_{-}}(u)>0$, then there is a unit vector $v\in T_{u}$ with $\ell_{\tau}'(u;v)<0$. In particular $u$ is neither a stationary point nor a local minimum of $\ell_{\tau}$.
\end{lemma}

\begin{proof}
Mass conservation \eqref{eq:mass} together with $\ell_{\tau}(u)=\sum_{k}\lvert\delta^{\tau}_{k}\rvert>0$ implies that $\Pc$ is nonempty, and on each cell of $\Pc$ we have $p^{\tau}_{k}>q^{\tau}_{k}\ge0$, hence $p^{\tau}_{k}>0$. Pick a surplus cell $k_{+}\in\Pc$; both $k_{\pm}$ lie in $\mathrm{supp}(u)$, so by the construction in the third part of Lemma~\ref{lem:dd} there is $w\in T_{u}$ realizing the zero-sum flow $\alpha_{k_{+}}=-\varepsilon$, $\alpha_{k_{-}}=+\varepsilon$ (with $\varepsilon>0$) and $0$ elsewhere. For this $w$ we have $dp^{\tau}_{k_{+}}(w)=-\varepsilon$, $dp^{\tau}_{k_{-}}(w)=+\varepsilon$ and $dp^{\tau}_{j}(w)=0$ on every other cell---on $\Pc\setminus\{k_{+}\}$, on $\Dc\setminus\{k_{-}\}$ and on all of $\Zc$. Taking the unit vector $v=w/\lVert w\rVert$ and using linearity of $dp^{\tau}_{k}$, substitution into \eqref{eq:dd} leaves only two of the three sums:
\[
\ell_{\tau}'(u;v)
=dp^{\tau}_{k_{+}}(v)-dp^{\tau}_{k_{-}}(v)
=\frac{(-\varepsilon)-(+\varepsilon)}{\lVert w\rVert}
=-\frac{2\varepsilon}{\lVert w\rVert}<0 .
\]
The amount removed from the surplus cell decreases $\delta^{\tau}_{k_{+}}>0$, and the amount poured into the deficient cell decreases $\lvert\delta^{\tau}_{k_{-}}\rvert$---the descent comes from both sides at once. Hence $u$ is not stationary, and since $\ell_{\tau}$ decreases strictly along $\gamma_{v}(t)$ for small $t>0$, it is not a local minimum either.
\end{proof}

\subsubsection*{Step 4: when the support condition holds}\label{app:support}

\begin{lemma}[Consequences of the support condition]\label{lem:supp}
Fix a slice $\tau$ and let $u\in\Sph$ satisfy $\ell_{\tau}(u)>0$ and the support condition. Then the following hold. First, $u$ is a stationary point of $\ell_{\tau}$. Second, mass conservation \eqref{eq:mass} gives
\[
\sum_{k\in\Pc}\delta^{\tau}_{k}=\sum_{k\in\Dc}q^{\tau}_{k}=\tfrac{1}{2}\,\ell_{\tau}(u)>0,
\]
every cell $k_{0}$ of $\Dc$ has $q^{\tau}_{k_{0}}>0$ (and $p^{\tau}_{k_{0}}=0$), and the surplus mass satisfies
\[
P_{\Pc}:=\sum_{k\in\Pc}p^{\tau}_{k}(u)\ \ge\ \tfrac{1}{2}\,\ell_{\tau}(u)\ >\ 0 .
\]
Third, for any deficient cell $k_{0}\in\Dc$ and any unit vector $v\in\ran\Pi^{\tau}_{k_{0}}$, along the great circle $u_{t}=\gamma_{v}(t)=\cos t\,u+\sin t\,v\in\Sph$ and in the range
\[
\sin^{2}t\ \le\ h_{0}:=\min\Bigl\{\min_{k\in\Pc}\frac{\delta^{\tau}_{k}}{p^{\tau}_{k}},\ q^{\tau}_{k_{0}}\Bigr\}\ >\ 0,
\]
the exact identity
\begin{equation}\label{eq:leak}
\ell_{\tau}(u_{t})=\ell_{\tau}(u)-2P_{\Pc}\,\sin^{2}t
\end{equation}
holds. In particular $u$ is not a local minimum of $\ell_{\tau}$.
\end{lemma}

\begin{proof}
We prove the second claim first. The support condition gives $p^{\tau}_{k}=0$ on $\Dc$, i.e.\ $\delta^{\tau}_{k}=p^{\tau}_{k}-q^{\tau}_{k}=-q^{\tau}_{k}$, and $\delta^{\tau}_{k}<0$ then forces $q^{\tau}_{k}>0$ for every $k\in\Dc$. Mass conservation \eqref{eq:mass} reads $\sum_{\Pc}\delta^{\tau}_{k}=-\sum_{\Dc}\delta^{\tau}_{k}=\sum_{\Dc}q^{\tau}_{k}$, so $\ell_{\tau}(u)=\sum_{\Pc}\delta^{\tau}_{k}+\sum_{\Dc}q^{\tau}_{k}=2\sum_{\Dc}q^{\tau}_{k}>0$. On each cell of $\Pc$ we have $p^{\tau}_{k}>q^{\tau}_{k}\ge0$, hence $p^{\tau}_{k}\ge\delta^{\tau}_{k}>0$, and summing gives $P_{\Pc}\ge\sum_{\Pc}\delta^{\tau}_{k}=\ell_{\tau}(u)/2>0$; in particular $h_{0}>0$.

Next the first claim. By the support condition $p^{\tau}_{k}=0$ on all of $\Dc$, and together with the cells of $\Zc\cap\{p^{\tau}_{k}=0\}$, the third part of Lemma~\ref{lem:dd} gives $dp^{\tau}_{k}\equiv0$ on those cells. The sum over $\Dc$ in \eqref{eq:dd} therefore vanishes and the sum over $\Zc$ reduces to the sum over $\Zc\cap\{p^{\tau}_{k}>0\}$ (the cells dropped contribute $0$; they are not ignored):
\[
\ell_{\tau}'(u;v)=\sum_{k\in\Pc}dp^{\tau}_{k}(v)+\sum_{k\in\Zc,\,p^{\tau}_{k}>0}\lvert dp^{\tau}_{k}(v)\rvert .
\]
For the same reason the first-order mass conservation $\sum_{k}dp^{\tau}_{k}(v)=0$ reads $\sum_{k\in\Pc}dp^{\tau}_{k}(v)=\allowbreak-\sum_{k\in\Zc,\,p^{\tau}_{k}>0}dp^{\tau}_{k}(v)$ on the surviving cells. Substituting,
\[
\ell_{\tau}'(u;v)
=\sum_{k\in\Zc,\,p^{\tau}_{k}>0}\bigl(\lvert dp^{\tau}_{k}(v)\rvert-dp^{\tau}_{k}(v)\bigr)\ \ge\ 0
\]
for every unit vector $v\in T_{u}$, i.e.\ $u$ is stationary. In words: by Lemma~\ref{lem:dd} the mass flows realizable at first order are the zero-sum flows on the support of $u$, but under the support condition that support consists of $\Pc$ and $\Zc\cap\{p^{\tau}_{k}>0\}$ only, i.e.\ of cells with nonnegative residual. Lowering the loss at first order would require pouring mass into a deficient cell, yet the deficient cells are empty and inflow into an empty cell begins only at second order in $t$ because $p^{\tau}_{k}=\lVert\Pi^{\tau}_{k}u\rVert^{2}$ is quadratic. Moving mass inside the support only trades a decrease at the source for an increase at the destination, with no first-order gain; that is the content of the inequality above.

Finally the third claim. Three basic facts. Since $k_{0}\in\Dc$, the support condition gives $p^{\tau}_{k_{0}}(u)=0$, and with $p^{\tau}_{k_{0}}(u)=\lVert\Pi^{\tau}_{k_{0}}u\rVert^{2}$ this means $\Pi^{\tau}_{k_{0}}u=0$. As $v\in\ran\Pi^{\tau}_{k_{0}}$ means $v=\Pi^{\tau}_{k_{0}}v$, self-adjointness gives $\langle u,v\rangle=\langle u,\Pi^{\tau}_{k_{0}}v\rangle=\langle\Pi^{\tau}_{k_{0}}u,v\rangle=0$ (complex orthogonality; in particular $v\in T_{u}$), and orthogonality of the projectors \eqref{eq:resolution} gives $\Pi^{\tau}_{k}v=\Pi^{\tau}_{k}\Pi^{\tau}_{k_{0}}v=0$ for $k\ne k_{0}$. Hence, first, the cross term vanishes and
\[
\lVert u_{t}\rVert^{2}
=\cos^{2}t\,\lVert u\rVert^{2}+2\cos t\,\sin t\,\mathrm{Re}\langle u,v\rangle+\sin^{2}t\,\lVert v\rVert^{2}
=\cos^{2}t+\sin^{2}t=1,
\]
so $u_{t}$ lies on $\Sph$. Second, applying $\Pi^{\tau}_{k}$ linearly to $u_{t}$,
\[
\Pi^{\tau}_{k}u_{t}=\cos t\,\Pi^{\tau}_{k}u+\sin t\,\Pi^{\tau}_{k}v
=\begin{cases}
\cos t\,\Pi^{\tau}_{k}u & (k\ne k_{0};\ \Pi^{\tau}_{k}v=0),\\[2pt]
\sin t\,v & (k=k_{0};\ \Pi^{\tau}_{k_{0}}u=0,\ \Pi^{\tau}_{k_{0}}v=v),
\end{cases}
\]
so that, applying $p^{\tau}_{k}(w)=\lVert\Pi^{\tau}_{k}w\rVert^{2}$ to each line and abbreviating $h=\sin^{2}t$, exactly
\[
p^{\tau}_{k}(u_{t})=\cos^{2}t\,\lVert\Pi^{\tau}_{k}u\rVert^{2}=(1-h)\,p^{\tau}_{k}(u)\ \ (k\ne k_{0}),
\qquad
p^{\tau}_{k_{0}}(u_{t})=\sin^{2}t\,\lVert v\rVert^{2}=h .
\]
Count the change of the absolute residuals in the range $h\le h_{0}$. For $k\in\Pc$ the new residual $\delta^{\tau}_{k}-h\,p^{\tau}_{k}$ stays nonnegative (because $h\le\delta^{\tau}_{k}/p^{\tau}_{k}$) and changes by $-h\,p^{\tau}_{k}$. For $k\in\Zc$ with $p^{\tau}_{k}>0$ (where $q^{\tau}_{k}=p^{\tau}_{k}$) the change is an increase $\lvert-h\,p^{\tau}_{k}\rvert=h\,p^{\tau}_{k}$. Cells of $\Zc$ with $p^{\tau}_{k}=0$ stay at $0$, and cells of $\Dc\setminus\{k_{0}\}$ are unchanged since $p$ remains $0$ there. At $k_{0}$ the residual is $\lvert h-q^{\tau}_{k_{0}}\rvert=q^{\tau}_{k_{0}}-h$ (because $h\le q^{\tau}_{k_{0}}$), a change of $-h$. Summing: the whole mass of $p$ sits on $\Pc$ and $\Zc\cap\{p^{\tau}_{k}>0\}$ (the cells of $\Dc$ and of $\Zc\cap\{p^{\tau}_{k}=0\}$ carry no mass), so writing $P_{\Zc_{+}}$ for the mass of the latter we have $P_{\Pc}+P_{\Zc_{+}}=1$ and
\[
\ell_{\tau}(u_{t})-\ell_{\tau}(u)=-h\,P_{\Pc}+h\,P_{\Zc_{+}}-h=-2h\,P_{\Pc},
\]
which is \eqref{eq:leak}. Since $P_{\Pc}>0$ by the second claim, $\ell_{\tau}(u_{t})<\ell_{\tau}(u)$ for $0<h\le h_{0}$, and $u_{t}\to u$ as $t\downarrow0$, so $u$ is not a local minimum.
\end{proof}

\subsubsection*{Proof of Theorem~\ref{thm:main-landscape}}\label{app:thm}

\begin{customthm}{\ref*{thm:main-landscape}}[Landscape theorem]
Let $L(u)>0$ and let the maximizing slice be unique. Then $u$ is a stationary point if and only if the support condition holds. Moreover, when the support condition holds, $u$ is not a local minimum: for any deficient cell $k_{0}\in\mathcal{D}$ and any unit vector $v\in\ran\Pi^{\tau_{\star}}_{k_{0}}$, along the great circle $\gamma_{v}$ and for all sufficiently small $t>0$,
\[
L(\gamma_{v}(t))\;=\;L(u)\;-\;2P_{\mathcal{P}}(u)\,\sin^{2}t,
\qquad
P_{\mathcal{P}}(u)=\sum_{k\in\mathcal{P}}p^{\tau_{\star}}_{k}(u)\;>\;0 .
\]
\end{customthm}

\begin{proof}
Write $\tau_{\star}$ for the maximizing slice and apply the four steps in order.

\emph{Step 1 (localization).} By Lemma~\ref{lem:localize}, the one-sided directional derivatives, stationarity and local minimality of $L$ at $u$ agree with those of $\ell_{\tau_{\star}}$; below we argue about $\ell_{\tau_{\star}}$ only. Note $\ell_{\tau_{\star}}(u)=L(u)>0$.

\emph{Step 2 (mass flows realizable at first order).} By Lemma~\ref{lem:dd}, $\ell_{\tau_{\star}}'(u;v)$ exists for every unit vector $v\in T_{u}$ and is given by \eqref{eq:dd}, and the flows realizable at first order form exactly the set \eqref{eq:flowset} of zero-sum flows on the support.

\emph{Step 3 (the support condition fails).} If it fails, Lemma~\ref{lem:violate} provides a unit vector $v$ with $\ell_{\tau_{\star}}'(u;v)<0$. Hence $u$ is not a stationary point of $\ell_{\tau_{\star}}$, and by Step 1 not one of $L$ either.

\emph{Step 4 (the support condition holds).} If it holds, the first claim of Lemma~\ref{lem:supp} makes $u$ a stationary point of $\ell_{\tau_{\star}}$, hence of $L$ by Step 1. Moreover, by the third claim of Lemma~\ref{lem:supp}, for any $k_{0}\in\Dc$ and any unit vector $v\in\ran\Pi^{\tau_{\star}}_{k_{0}}$ we have $\ell_{\tau_{\star}}(\gamma_{v}(t))=\ell_{\tau_{\star}}(u)-2P_{\Pc}\sin^{2}t$ in the range $\sin^{2}t\le h_{0}$. Since $\lVert\gamma_{v}(t)-u\rVert\to0$ as $t\downarrow0$, for sufficiently small $t>0$ the point $\gamma_{v}(t)$ lies in the neighbourhood of Lemma~\ref{lem:localize} (the ball of radius $\Delta/4$), so that $L(\gamma_{v}(t))=\ell_{\tau_{\star}}(\gamma_{v}(t))$, and $\sin^{2}t\le h_{0}$ holds as well; the claimed identity therefore holds for $L$. By the second claim of Lemma~\ref{lem:supp}, $P_{\Pc}=P_{\mathcal{P}}(u)\ge\ell(u)/2>0$, so the identity gives a strict descent of $L$ and $u$ is not a local minimum of $L$.

Combining: the support condition implies stationarity (Step 4) and its failure implies non-stationarity (Step 3), so stationarity is equivalent to the support condition; and when it holds, $u$ is not a local minimum, as shown in Step 4.
\end{proof}

If, in addition, $p^{\tau_{\star}}_{k}(u)>0$ on every cell of $\tau_{\star}$ (interior condition), then the nonempty $\Dc$ has mass on all of its cells, so the support condition cannot hold and $u$ is not even a stationary point.

\subsubsection*{Proof of Corollary~\ref{cor:main-ties}}\label{app:cor}

\begin{customcor}{\ref*{cor:main-ties}}
Every spurious local minimum of $L$ (a local minimum with positive value) lies in the \emph{tie set}, the set of points at which several slices share the maximum. Moreover, under Assumption~\ref{as:meas} and with exact data $q^{\tau}=p^{\tau}(u_{\star})$, for almost every truth $u_{\star}$ the global minimum of $L$ is the truth up to a global phase.
\end{customcor}

\begin{proof}
The first statement is immediate from Theorem~\ref{thm:main-landscape}: a point with $L(u)>0$ at which the maximizing slice is unique is not a local minimum, whether the support condition holds or fails, so every local minimum with $L>0$ is a point at which the maximizing slice is not unique, i.e.\ lies in the tie set.

For the second statement, $L(u)=0$ is by the definition \eqref{eq:loss} equivalent to $p^{\tau}(u)=q^{\tau}$ for all $\tau$, which under exact data $q^{\tau}=p^{\tau}(u_{\star})$ reads $u^{\dagger}\Pi^{\tau}_{k}u=u_{\star}^{\dagger}\Pi^{\tau}_{k}u_{\star}$ for all $\tau$ and $k$. By the substring expansion \eqref{eq:substring}, each cell projector $\Pi^{\tau}_{k}$ of a setting $\tau=(S,b)$ is a linear combination of the Pauli substrings $P_{b,T}$ ($T\subseteq S$) belonging to that slice, and conversely each $P_{b,T}$ is a signed sum of the $\{\Pi^{\tau}_{k}\}_{k}$; hence agreement of the cell probabilities is equivalent to $\Tr(P\,uu^{\dagger})=\Tr(P\,u_{\star}u_{\star}^{\dagger})$ for every Pauli string $P$ belonging to a slice. Under Assumption~\ref{as:meas} the settings $\tau=(S,b)$ run over all blocks with $\lvert S\rvert=N$ and all labels $b\in\{x,y,z\}^{S}$, so every Pauli string of weight at most $N$ belongs to some slice; therefore $L(u)=0$ is equivalent to the agreement of all Pauli expectations of weight at most $N$, i.e.\ of all $N$-body reduced density matrices of the two states. If the truth $u_{\star}$ is determined uniquely, up to a global phase, among pure states by these expectations, then $L(u)=0$ implies $u=e^{\mathrm{i}\theta}u_{\star}$; since $L\ge0$ and $L(u_{\star})=0$, the global minimizers of $L$ are exactly $e^{\mathrm{i}\theta}u_{\star}$. That almost every pure state has this property when $n=2N$ is Proposition~\ref{prop:udp}.
\end{proof}

\end{document}